\documentclass[aps, prx, reprint, superscriptaddress, notitlepage, floatfix, nofootinbib]{revtex4-2}

\usepackage[utf8]{inputenc}
\usepackage{amsmath,amssymb,amsthm,amscd,latexsym,epsfig,bbm}

\usepackage{physics}
\usepackage{dsfont}
\usepackage{graphicx}
\usepackage{xcolor}
\usepackage{graphicx}
\usepackage[export]{adjustbox}

\usepackage{tensor}

\usepackage{cancel}
\usepackage{color}
\usepackage{tikz}
\usepackage{mathrsfs}
\usepackage{relsize}
\usepackage[linktocpage]{hyperref}
\usepackage[all]{xy}
\hypersetup{colorlinks=true,citecolor=blue,linkcolor=blue, urlcolor=blue, breaklinks=true}
\usepackage{verbatim}
\usepackage{braket}
\usepackage{tikz-cd}

\usepackage{bm} 
\usepackage{subfigure} 
\usepackage{environ}
\usepackage{url}
\usepackage[margin=1in]{geometry}
\usepackage[export]{adjustbox}
\usepackage{float}

\usepackage{framed}
\usepackage{booktabs}
\usepackage{longtable}
\usepackage{amstext}
\usepackage{array}
\newcolumntype{L}{>{$}l<{$}} 
\newcolumntype{C}{>{$}c<{$}} 

\def\beq{\begin{equation}}
\def\eeq{\end{equation}}
\def\be{\begin{equation}}
\def\ee{\end{equation}}

\def\ket#1{\vert #1 \rangle}
\def\bra#1{\langle #1 \vert}
\def\ev#1{\langle #1 \rangle}

\def\me#1#2#3{\langle #1 \vert #2 \vert #3 \rangle}

\newcommand{\z}{\mathbb{Z}}

\def\f2{{\mathbb F}_2}

\theoremstyle{plain}
\theoremstyle{plain}

\providecommand{\theoremname}{Theorem}
\providecommand{\theoremtextname}{Theorem}

\theoremstyle{plain}
\providecommand{\propositionname}{Proposition}

\begin{document}

\title{Generating logical magic states with the aid of non-Abelian topological order}

\author{Sheng-Jie Huang}
\affiliation{Mathematical Institute, University of Oxford, Woodstock Road, Oxford, OX2 6GG, United Kingdom}
\email{sheng-jie.huang@maths.ox.ac.uk}

\author{Yanzhu Chen}
\affiliation{Department of Physics, Florida State University, Tallahassee, FL 32306, USA}
\email{yanzhu.chen@fsu.edu}

\begin{abstract} 

In fault-tolerant quantum computing with the surface code, non-Clifford gates are crucial for universal computation. However, implementing these gates using methods like magic state distillation and code switching requires significant resources.  
In this work, we propose a new protocol that combines magic state preparation and code transformation to realize logical non-Clifford operations with the potential for fault tolerance. Our approach begins with a special logical state in the $\mathbb{Z}_4$ surface code. By applying a sequence of transformations, the system goes through different topological codes, including the non-Abelian $D_4$ quantum double model. 
This process ultimately produces a magic state encoded in the $\mathbb{Z}_{2}$ surface code. A logical $T$ gate can be implemented in the standard $\mathbb{Z}_2$ surface code by gate teleportation. 
In our analysis, we employ a framework where the topological codes are represented by their topological orders and all the transformations are considered as topological manipulations such as gauging symmetries and condensing anyons. This perspective is particularly useful for understanding transformations between topological codes.  

\end{abstract}

\maketitle

\tableofcontents


\section{Introduction}

Quantum computation has promised revolutionary capabilities surpassing classical computation, if we can efficiently deal with the inevitable errors corrupting the quantum information. Various quantum error correction (QEC) codes have been designed to encode logical information in a code subspace, which is protected by frequent checks and corrections on the physical qubits~\cite{Shor1995, Kitaev1997, Cochrane1999, Gottesman2001, Gottesman2009}. Among the most promising QEC codes is the surface code~\cite{Dennis2002}, celebrated for its high error threshold and simple planar connectivity requirements. These features make it a leading candidate for fault-tolerant quantum computation~\cite{Fowler2012,Horsman2012,Fowler2018,Litinski2019gameofsurfacecodes}. To realize universal fault-tolerant quantum computing, a crucial task is engineering a universal set of logical quantum gates in the code subspace. Transversal gates, which can be constructed from constant-depth local-unitary circuits on physical qubits, are fault-tolerant since they do not propagate errors throughout the entire code block as the code block scales up. However, for any given code transversal gates alone cannot form a universal set of logical gates~\cite{Eastin2009}. In the surface code, transversal gates are limited to logical Clifford gates~\cite{Brown2017,Fowler2018,Litinski2019gameofsurfacecodes}. Achieving universality requires at least one logical non-Clifford gate, such as the $T$ gate. 

The need to construct a logical non-Clifford gate presents a critical bottleneck for fault-tolerant quantum computation. A leading proposal for the $T$ gate in the surface code is magic state injection and distillation~\cite{Bravyi2005magicstate,Fowler2012,Gidney2019efficientmagicstate,Litinski2019magicstate}. It involves extra measurements on sufficiently many copies of noisy magic states, which introduces a large resource overhead. The approach of code switching seeks to temporarily transform the QEC code to another code in which a non-Clifford gate can be implemented as a transversal gate, whenever the need for non-Clifford gates arises~\cite{Anderson2014,Kubica2015,Bombin2015,Daguerre2024}. It requires no state distillation but still a measurement overhead associated with error correction. Introducing non-Clifford operations by switching between topological codes will be particularly desirable, due to their robustness against local errors and their high error thresholds. However, it has been shown that switching between 2-D and 3-D color codes does not improve substantially on the resource overhead compared to magic state distillation while having a lower error threshold~\cite{Beverland2021}.

Similar to the idea of code switching, one may introduce non-Clifford operations by exploiting an intermediate higher-dimensional Hilbert space. Attempts were made to leverage generalized $\z_4$ Clifford operations on a $4$-dimensional qudit to complement the $\z_2$ Clifford operations~\cite{Moussa2016,Moussa2016Fold}. Interpreted at the logical level, this procedure offers a code switching method. However, in this approach a special resource state on the $4$-dimensional qudit is essential to forming a universal set of gates. Fault-tolerant preparation of this state at the logical level also requires resource-intensive distillation.

In this work, we propose a novel method to introduce non-Clifford gates in topological codes by combining magic state preparation and code transformation. Specifically, we prepare the logical magic state in a topological code by going through an intermediate 2-D non-Abelian topological code. This method circumvents the state distillation process. 
With the purpose being magic state preparation, we only need to transform one specific logical state, instead of any arbitrary logical state during computation, through different codes, which is different from the usual code switching methods. At the heart of our method is the intermediate non-Abelian topological code. It falls within a broad class given by Kitaev’s quantum double models~\cite{Kitaev2003}, of which the surface code is a special example. Defined for a finite group $G$, the $G$ quantum double model is constructed on a lattice of $|G|$-dimensional qudits. In particular, we make use of the $D_4$ quantum double model. 

Topological codes can be considered as models on a lattice of qubits or qudits exhibiting topological orders, with gapped boundary conditions specified. The topological order is described by the algebraic theory of anyons, known as unitary modular tensor categories~\cite{Nayak2008,wang2008}, which provides a continuum description of the code. The anyon theories with gapped boundary conditions form the so-called \emph{Drinfeld center} $\mathcal{Z}(\mathcal{C})$ of some input fusion category $\mathcal{C}$~\cite{Drinfeld2010}. 
The $G$ quantum double model realizes the topological order $\mathcal{Z}(\text{Rep}(G))$, where $\text{Rep}(G)$ is the category of the representation of $G$. When $G$ is a non-Abelian group, the model hosts non-Abelian anyons, bringing new possibilities in quantum computation.

Representing the QEC codes involved in our method with the topological orders they realize, the operations in the entire procedure can be described completely using the language of topological manipulations, such as gauging symmetries~\cite{Haegeman2015,Kubica2018,Tantivasadakarn2024,Tantivasadakarn2023Long} and condensing anyons~\cite{Bais2009,Eliens2014,Kong2014,Cong2017PRL,Cong2017,Burnell2018,Lou2021,Kesselring2024}. Switching between different topological codes, a logical state that is challenging to prepare in one code can be transformed from a logical state easily prepared in another. Fixing the initial, intermediate, and final topological codes, one can translate the topological manipulations back to operations on the lattice.

Specifically, we start from the $\z_4$ surface code, which is a generalization of the standard surface code on physical $4$-dimensional qudits. A special state in this code is initially prepared using logical generalized $\z_4$ Clifford gates. We gauge the charge conjugation symmetry to obtain a code that realizes the topological order of the $D_{4}$ quantum double model.  
Then, an anyon condensation procedure is performed to obtain the $\z_{2}^2$ surface code. Treating this as two logical qubits, we can extract one logical magic state through transversal gates. Alternatively, a further anyon condensation step can be performed to obtain a code that realizes the same topological order as the standard surface code.
We refer to this code as the condensed $\z_{2}$ surface code. Through this procedure, the initial state is transformed into the logical magic state in the condensed $\z_{2}$ surface code while the code distance is preserved. We also provide the local gates that can turn this into the standard $\z_{2}$ surface code. 
The magic state prepared this way can be subsequently used for implementing a $T$ gate in the surface code through gate teleportation~\cite{Bravyi2005magicstate,Fowler2012,Gidney2019efficientmagicstate,Litinski2019magicstate}. On the lattice, these topological manipulations are implemented by using adaptive finite-depth local-unitary circuits~\cite{Gottesman1999,Piroli2021,Verresen2021,Tantivasadakarn2024,Bravyi2022,Tantivasadakarn2023nonabelian,Tantivasadakarn2023Long,Li2023,Williamson2024}, which combine local unitary gates, single-site measurements, global classical communication, and local unitary feed-forward operations. 

\begin{figure}
	\centering
	\includegraphics[width=0.4\textwidth]{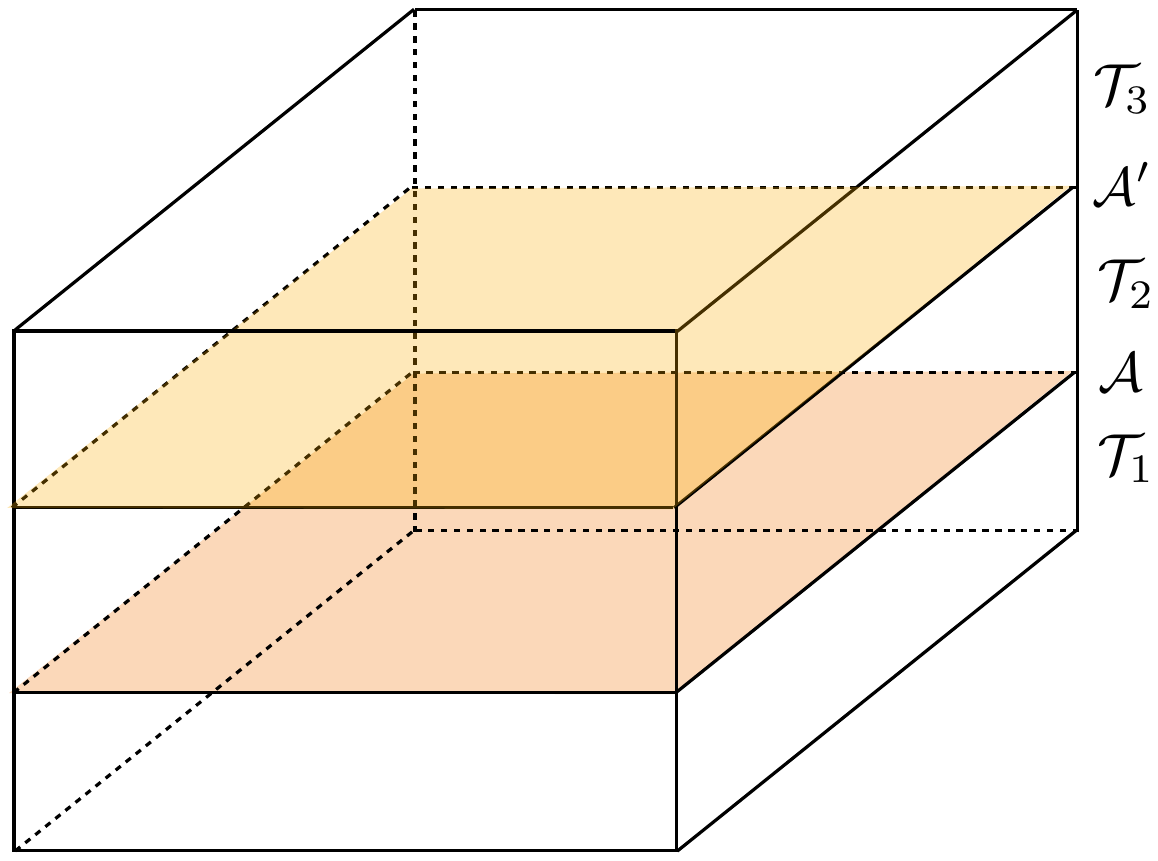}
	\caption{The `sandwich' structure representing our code transformation procedure, which can analyzed categorically in the algebraic theory of anyons. Starting from a code realizing the topological order $\mathcal{T}_{1}$, we perform a set of topological manipulations, including gauging symmetries and condensing anyons, to transform the logical information into the final code with $\mathcal{T}_{3}$ order. Each topological manipulation is described by a gapped interface between the topological orders.}
	\label{fig:sandwich_abs}
\end{figure}

In addition to a novel magic state preparation method, our method provides an abstract framework of studying transformations between topological codes, under which the transformation on the logical information can be analyzed without details about the lattice models. 
Under this framework, each topological manipulation between two topological orders is described by a gapped interface. The topological orders together with the designed set of the topological manipulations form a `sandwich' structure, illustrated in Fig.~\ref{fig:sandwich_abs}. With the topological order $\mathcal{T}_{1}$ in the initial code and $\mathcal{T}_{3}$ in the final code fixed, the transformation of the logical information is determined by the choices of the intermediate topological order $\mathcal{T}_{2}$ and the gapped interfaces $\mathcal{A}$ and $\mathcal{A}'$.

Non-Abelian topological orders have found applications in quantum computation through the encoding of information in the fusion space of non-Abelian anyons~\cite{Mochon2004,Nayak2008,Cui2015}.  
Hole encoding has also been studied in non-Abelian topological codes~\cite{Cong2017PRL,Cong2017}. With hole encoding, small islands of the non-Abelian $S_3$ quantum double model in the $\z_2$ surface code can be employed to generate non-stabilizer states~\cite{Laubscher2019}. Code transformation through anyon condensation has been explored for quantum double models with fusion space encoding and for Floquet codes~\cite{Ren2023}.

This manuscript is organized as follows. In the next section, we review Kitaev's quantum double model for a finite group $G$. We first give a special example with the group $G = \z_{n}$ in Sec.~\ref{sec:zn_surface_code}. With a choice of boundary conditions, this defines the $\z_n$ surface code. We review logical Clifford operations in the $\z_{4}$ surface code, which will be used in our method. Then, we describe the general definition of quantum double models in Sec.~\ref{sec:quantum_double}, including a discussion of excitations and the operators producing them. With the definition in mind, we show the explicit form of the $G = D_{4}$ quantum double model (or $D_4$ surface code, if boundary conditions are specified) in Sec.~\ref{sec:D4_model}.   
In Sec.~\ref{sec:magic_cat}, we introduce our main idea to generate the magic state using the continuum description of the initial, intermediate, and final codes. 
We then translate the topological manipulations to operations on the lattice in Sec.~\ref{sec:magic_lattice}, each manipulation followed by a discussion of how the logical state transforms. 
In Sec.~\ref{sec:gauge_c}, we describe how to gauge the charge conjugation symmetry in the $\z_{4}$ surface code to obtain the $D_{4}$ surface code.  
In Sec.~\ref{sec:eg_cond}, we describe the anyon condensation procedure that reaches an intermediate model equivalent to the $\z_{2}^{2}$ surface code, from which one magic state can be extracted by transversal gates (Sec.~\ref{sec:disentangle_tc}) or further anyon condensation (Sec.~\ref{sec:condensed_z2_surface_code}).
In Sec.~\ref{sec:tgate}, we show how to perform the $T$ gate in the standard $\z_{2}$ surface code by consuming the magic state prepared in the condensed $\z_{2}$ surface code through gate teleportation. 
In Sec.~\ref{sec:discussion}, we summarize our result and discuss the implications and the questions we leave open. 
In Appendix~\ref{app:D4_model_ribbon}, we provide details of some ribbon operators in the $D_{4}$ surface code that will be used in this work. The basic idea of anyon condensation and a description of the gapped interfaces is provided in Appendix~\ref{app:anyon_condensation}. In Appendix~\ref{app:z4z2}, we show that performing direct anyon condensation to transform from a $\z_{4}$ surface code to a $\z_{2}$ surface code can not achieve the desired transformation on the logical information. In Appendix~\ref{app:bdy_cz2}, we give details of the boundary Hamiltonian terms in the condensed $\z_{2}$ surface code.

\section{Review: Kitaev's quantum double models}
\label{sec:review_Kitaev}

In this section, we will first review the $\z_{n}$ surface code, which is the $\z_{n}$ toric code with open boundary conditions and a generalization of the standard $\z_2$ surface code. With this as a special example, we will review Kitaev's quantum double models defined for a given finite group $G$. Throughout our procedure of preparing the magic state, we will make use of the $G=\z_2$, $G=\z_4$, and $G=D_4$ cases, which correspond to the standard $\z_2$ surface code, the $\z_{4}$ surface code, and the $D_4$ quantum double model.

\subsection{$\z_{n}$ surface code}
\label{sec:zn_surface_code}

The $\z_{n}$ surface code is the $\z_{n}$ toric code with open boundary conditions, which we will specify later. In the $\z_{n}$ toric code, the physical Hilbert space at each edge of the square lattice is $n$-dimensional, referred to as an $n$-dimensional qudit. The vertical and horizontal edges are oriented upward and rightward, respectively. The orientations are important when discussing the string operators on an oriented path that create excitations. In the $n=2$ case, both directions of a path are equivalent so the lattice edges do not need to be oriented. 
The generalized Pauli-X and Pauli-Z operators for the qudit are defined as 
\begin{equation}
    \tilde{X} = \sum_{j \in \z_{n}} \ket{j+1}\bra{j}, \quad \tilde{Z} = \sum_{j \in \z_{n}} \omega^{j} \ket{j}\bra{j},
\end{equation}
where $\omega = e^{i 2\pi/n}$. These operators satisfy the relations
\begin{equation}
    \tilde{X}^{n} = 1, \quad \tilde{Z}^{n} = 1
\end{equation}
and the commutation relation
\begin{equation}
    \tilde{Z}\tilde{X} = \omega \tilde{X}\tilde{Z}. 
\end{equation}
The Hamiltonian of the $\z_{n}$ toric code is a stabilizer Hamiltonian, given by terms $A_v$ associated with vertices $v$ and $B_p$ associated with plaquettes $p$
\begin{equation}
    H = -\sum_{v} A_{v} - \sum_{p} B_{p},
\label{eq:zntoricH}
\end{equation}
where $A_{v} = \frac{1}{n} \sum_{j=0}^{n-1} (A_{v}^{(r)})^{j}$, $B_{p} = \frac{1}{n} \sum_{j=0}^{n-1} (B_{p}^{(r)})^{j}$, and
\begin{equation}
    A^{(r)}_{v} = \includegraphics[width=.25\linewidth,valign=c]{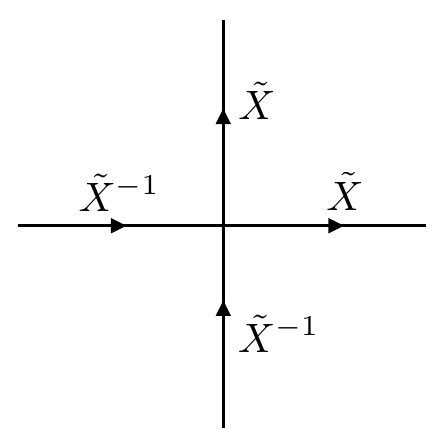}, \quad B^{(r)}_{p} = \includegraphics[width=.23\linewidth,valign=c]{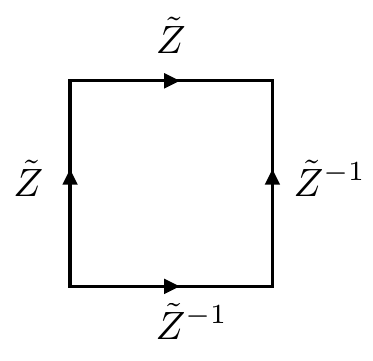}.
\end{equation}
The $\z_{n}$ toric code realizes the $\z_{n}$ toric code (TC) topological order on a torus, which we denote as $\mathcal{Z}(\z_{2})$. The bulk excitations are generated by anyons $e$ and $m$. The fusion rules coincide with the multiplication rules of the $\z_{n} \times \z_{n}$ group: $e^{n} = 1$ and $m^{n} =1$. The self-statistics of an arbitrary anyon $e^{p}m^{q}$  (with $p,q \in \z_{n}$) is given by
\begin{equation}
    \theta(e^{p}m^{q}) = \omega^{pq}. 
\end{equation}
The mutual braiding statistics between $e$ and $m$ is given by
\begin{equation}
    B_{\theta}(e,m) = \omega. 
\end{equation}

We now discuss the anyon excitations of the $\z_{n}$ toric code. The anyonic excitations are created by string operators, which can be organized into $e$-type and $m$-type. The $e$-type string operator $W^{(e)}_{\gamma}$ on a directed path $\gamma$ can be decomposed into products of short string operators: $W^{(e)}_{\gamma} = \prod_{l\in \gamma} W_{l}^{(e)}$, where the short string operators on one edge along the path can be represented pictorially as
\begin{equation}
    W^{(e)}_{l} = \includegraphics[width=.2\linewidth,valign=c]{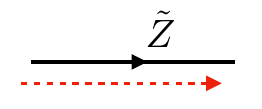}, \includegraphics[width=.2\linewidth,valign=c]{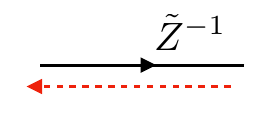},  \includegraphics[width=.09\linewidth,valign=c]{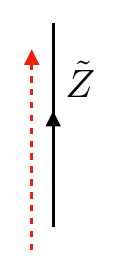}, \includegraphics[width=.12\linewidth,valign=c]{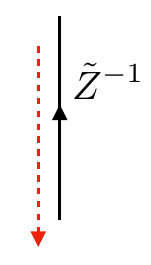}.
\end{equation}
Here the dashed red lines denote the orientation of the path. One can see that when $n=2$, there is no need to assign an orientation to the lattice edges. An open $e$-string operator $W^{(e)}_{\gamma}$ fails to commute with the vertex terms $A_{v}$ of the stabilizer Hamiltonian at the endpoints of the path $\gamma$. Specifically, at the initial vertex $v_{i}$ of the path $\gamma$, we have $A_{v_{i}} W^{(e)}_{\gamma} = \omega W^{(e)}_{\gamma} A_{v_{i}}$, and, at the final vertex $v_{f}$, we have $A_{v_{f}} W^{(e)}_{\gamma} = \omega^{n-1} W^{(e)}_{\gamma} A_{v_{f}}$. This means that the open $e$-string operator $W^{(e)}_{\gamma}$ creates an $e$ particle at $v_{i}$ and an $e^{n-1}$ particle at $v_{f}$. 

Similarly, the $m$-type string operator $W^{(m)}_{\gamma}$ on a directed path $\gamma$ can be decomposed into products of short string operators: $W^{(m)}_{\gamma} = \prod_{l\in \gamma} W_{l}^{(m)}$, where the short string operators on one edge along the path can be represented pictorially as
\begin{equation}
    W^{(m)}_{l} = \includegraphics[width=.16\linewidth,valign=c]{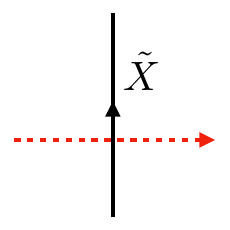}, \includegraphics[width=.16\linewidth,valign=c]{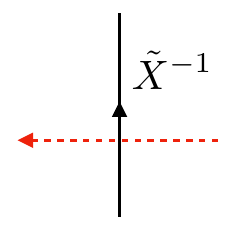},  \includegraphics[width=.16\linewidth,valign=c]{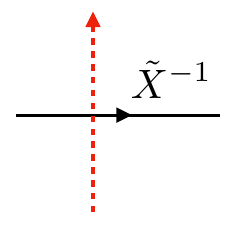},  \includegraphics[width=.16\linewidth,valign=c]{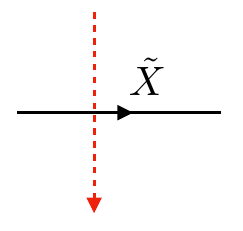}.
\end{equation}
An open $m$-string operator $W^{(m)}_{\gamma}$ fails to commute with the plaquette terms $B_{p}$ of the stabilizer Hamiltonian at the endpoints of the path $\gamma$. 
Specifically, at the initial plaquette $p_{i}$ of the path $\gamma$, we have $B_{p_{i}} W^{(m)}_{\gamma} = \omega W^{(m)}_{\gamma} B_{p_{i}}$, and, at the final vertex $p_{f}$, we have $B_{p_{f}} W^{(m)}_{\gamma} = \omega^{n-1} W^{(m)}_{\gamma} B_{p_{f}}$. Therefore, the open $m$-string operator $W^{(m)}_{\gamma}$ creates an $m$ particle at $p_{i}$ and an $m^{n-1}$ particle at $p_{f}$. 

The $\z_{n}$ surface code is defined on a square lattice with alternating smooth and rough boundary conditions. We choose the left and right boundaries to be smooth, and the top and bottom boundaries to be rough. This means that in addition to the toric code Hamiltonian terms Eq.~\eqref{eq:zntoricH} in the bulk, there are additional boundary terms, which can be represented graphically as:
\begin{align}
    A_{v}^{\rm L} &= \includegraphics[width=.2\linewidth,valign=c]{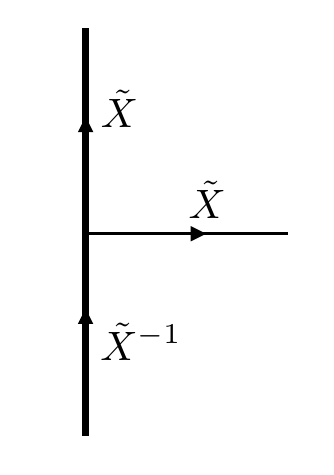}, \quad  A_{v}^{\rm R} = \includegraphics[width=.2\linewidth,valign=c]{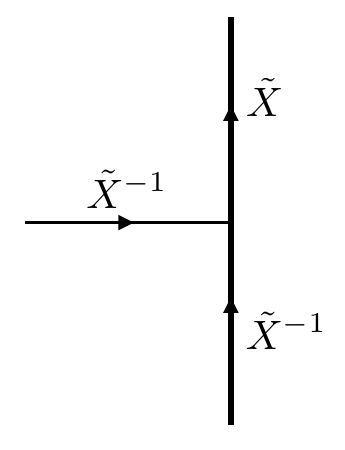}
\end{align}
for the left and right boundaries, and
\begin{align}
    B_{p}^{\rm T} &=\includegraphics[width=.22\linewidth,valign=c]{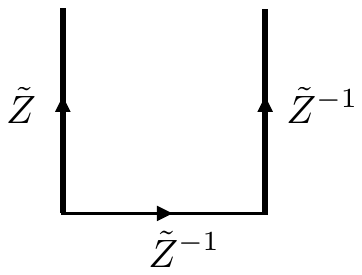} , \quad B_{p}^{\rm B} = \includegraphics[width=.22\linewidth,valign=c]{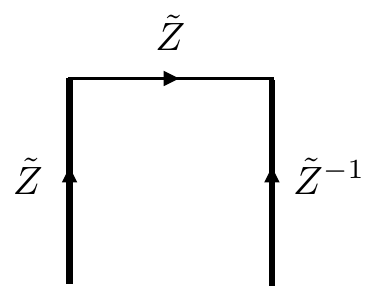}
\end{align}
for the top and bottom boundaries, where the boundaries are indicated by thick lines. The Hamiltonian terms commute with one another and generate a stabilizer group $\mathcal{S}_{SC}$:
\begin{equation}
    \mathcal{S}_{SC} \equiv \ev{\{A_{v}\},\{B_{p}\},\{A_{v}^{\rm L}\},\{A_{v}^{\rm R}\},\{B_{p}^{\rm T}\},\{B_{p}^{\rm B}\}}.
\label{eq:ssc}
\end{equation}
The logical subspace for the $\z_{n}$ surface code, which is the mutual $+1$ eigenspace of the stabilizers, is $n$-dimensional. 

The boundary stabilizers of the rough boundaries (top and bottom) commute with the $e^{k}$-string operators for $k=0,1,...,n-1$ that terminate at the boundary. We say that the $e^{k}$ particles condense on the rough boundaries. The $e^{k}$-string operators that connect the two opposite rough boundaries form the logical $\bar{Z}^{k}$ operators. Similarly, the boundary stabilizers of the smooth boundaries (left and right) commute with the $m^{k}$-string operators that terminate at the boundary. We say that the $m^{k}$ particles condense on the smooth boundaries. The $m^{k}$-string operators that connect the two opposite smooth boundaries form the logical $\bar{X}^{k}$ operators.

\subsubsection{Clifford operations in the $\z_{4}$ surface code}
\label{sec:z4clifford}

In this work, we are only concerned with the $n=2, 4$ cases of the $\z_n$ surface code. The former is just the standard qubit case. We now review the Clifford operations for the $4$-dimensional qudit and briefly discuss how to implement the logical Clifford operations in the $\z_{4}$ surface code. Similar to the qubit case, the single-qudit Clifford group can be generated up to a phase~\cite{Farinholt2014,Moussa2016} by a discrete Fourier transform
\begin{equation}
    \tilde{H} \ket{j} = \sum_{k \in \z_{4}} \frac{i^{jk}}{2} \ket{k},
\label{eq:z4_Fourier}
\end{equation}
and
the phase gate
\begin{equation}
    \tilde{S} \ket{k} = e^{\frac{i \pi k^{2}}{4}} \ket{k}.
\label{eq:z4_S}
\end{equation}
The multi-qudit Clifford group can be generated by the single-qudit Clifford group on each qudit and a $\widetilde{CX}$ gate between all pairs of qudits
\begin{equation}
    \widetilde{CX} \ket{j}\ket{k} = \ket{j}\ket{k+j}.
\end{equation}

The logical Pauli $\bar{Z}$ and $\bar{X}$ operations are realized by $e$-strings and $m$-strings connecting opposite boundaries, respectively, as discussed above. The logical Fourier transform $\bar{H}$ and the logical phase gate $\bar{S}$ are fold-transversal gates as shown in Ref.~\cite{Moussa2016Fold}, where the operation of folding along a diagonal of the square lattice is available. 
If the folding operation is not available, the $\bar{H}$ and $\bar{S}$ gates can still be implemented fault tolerantly. The $\bar{H}$ gate can be realized by a physical $\tilde{H}$ on each qudit, followed by a $90$-degree rotation of the surface code patch, a standard operation in lattice surgery~\cite{Cowtan2022}. We expect that the construction of the logical $S$ gate in Ref.~\cite{Gidney2024Ybasis} through moving the twist defects diagonally in the standard qubit surface code can be generalized to realizing the $\bar{S}$ gate in the qudit surface code. 
The logical $\overline{CX}$ gate can be implemented by the standard merge-split procedure in lattice surgery~\cite{Cowtan2022}. 

\subsection{Kitaev's quantum double models}
\label{sec:quantum_double}

Having seen a special case, the $\z_n$ surface code, we now review Kitaev's quantum double $D(G)$ model for a general finite group $G$~\cite{Kitaev2003}. Given a square lattice, we place a qudit at each edge, whose Hilbert space is $\mathbb{C}(G)$ with an orthonormal basis labeled by the group elements $\{ \ket{g} : g \in G \}$. The vertical and horizontal edges are oriented upward and rightward, respectively. 
The qudit at an edge $l$ is equipped with left and right multiplication operators associated with group elements $g\in G$
\begin{equation}
    L^{g}(l) \ket{h}_l = \ket{gh}_l, \quad R^{g}(l) \ket{h}_l = \ket{hg^{-1}}_l,
\end{equation}
as well as projection operators
\begin{equation}
    T^{g}_{+}(l) \ket{h}_l = \delta_{g,h} \ket{h}_l, \quad T^{g}_{-}(l) \ket{h}_l = \delta_{g^{-1},h} \ket{h}_l.
\end{equation}
For $g\in G$, at each vertex $v$ and plaquette $p$, we define the operators
\begin{equation}
    A^{(g)}_{v} = \includegraphics[width=.24\linewidth,valign=c]{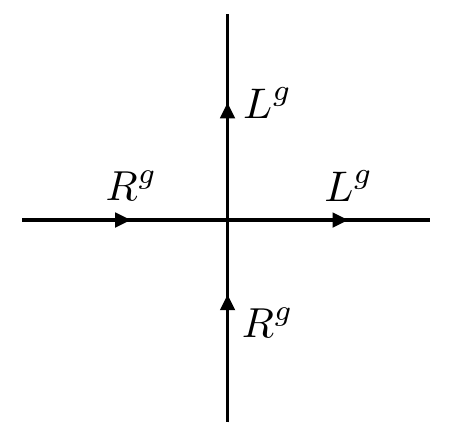},
\end{equation}
\begin{equation}
    B^{(g)}_{p} = \sum_{g_{1},g_{2},g_{3},g_{4} \in G} \delta_{g,g_{1}g_{2}g_{3}^{-1}g_{4}^{-1}} \Biggr|\includegraphics[width=.14\linewidth,valign=c]{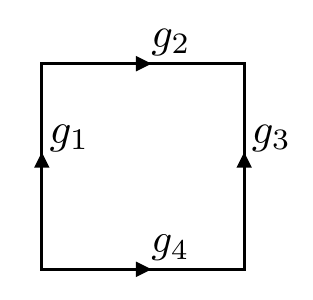} \Biggr\rangle \Biggr\langle \includegraphics[width=.14\linewidth,valign=c]{QDBP.pdf} \Biggr|.
\end{equation}
The quantum double Hamiltonian is given by
\begin{equation}
    H_{D(G)} = - \sum_{v} A_{v} - \sum_{p} B_{p},
\label{eq:qdoubleH}
\end{equation}
where the summations are over all vertices $v$ and plaquettes $p$, respectively, and the $A_{v}$ and $B_{p}$ terms are defined as
\begin{equation}
    A_{v} = \frac{1}{|G|} \sum_{g\in G} A_{v}^{(g)}, \quad  B_{p} = B_{p}^{({\rm Id})}.
\end{equation}
Since the vertex and plaquette operators $A_{v}$ and $B_{p}$ all commute and are projectors, the ground space of $H_{D(G)}$ is the simultaneous eigenspace of these operators with eigenvalues equal to $1$.

The excited states can be described in terms of localized particle excitations associated with sites $s$, which are formed by a plaquette $p$ and one of its vertices $v$. The particle excitations (or anyons) in the quantum double model can be represented in the form of $a=([g],\pi_{g})$, where $g \in G$ is a group element, $[g]$ denotes the conjugacy class $[g] = \{ k g k^{-1} | k \in G \}$, and $\pi_{g} \in \text{Rep}(C_{g})$ is the irreducible representation of the centralizer $C_{g}$ of $g$. We will sometimes call a particle $([g],I_0)$ a pure magnetic flux, where $I_0$ is a trivial representation, and we call $([{\rm Id}],\pi_{1})$ a pure electric charge, where $\pi_{1}$ is an irreducible representation of the entire group $G$.

\begin{figure}
	\centering
	\includegraphics[width=0.25\textwidth]{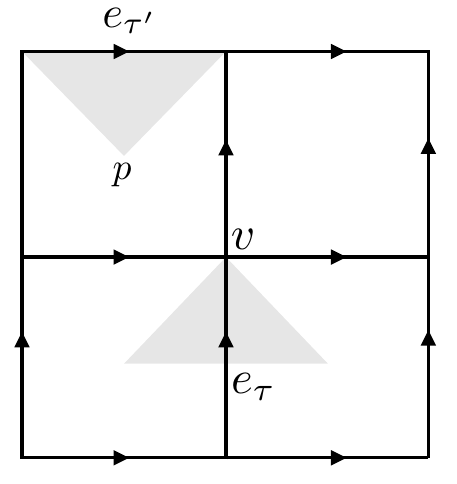}
	\caption{A direct triangle is one with the long edge aligned a lattice edge $e_{\tau'}$ and the opposite vertex at the center of a plaquette $p$. A dual triangle is one with the long edge intersecting a lattice edge $e_{\tau}$ and the opposite vertex at a vertex $v$.}
	\label{fig:triangle}
\end{figure}

The excitations and the logical operators in the quantum double model are formed by ribbon operators~\cite{Bombin2008,Cong2017}. A ribbon consists of a sequence of sites connecting a starting site $s_{0}=(v_{0},p_{0})$ to an ending site $s_{1}=(v_{1},p_{1})$ by adjoining the direct and dual triangles along the path. As illustrated in Fig.~\ref{fig:triangle}, a direct triangle is one with the long edge aligned with a lattice edge and the opposite vertex at the center of a plaquette $p$. A dual triangle is one with the long edge aligned with an edge of the dual lattice and the opposite vertex at a vertex $v$. 
For each pair $(h,g) \in G$, we define the basic ribbon operators for a dual triangle $\tau$ and a direct triangle $\tau'$, respectively,
\begin{equation}
    F_{\tau}^{h,g} = \delta_{{\rm Id},g} L_{\tau}^{h}, \quad F_{\tau'}^{h,g} = T_{\tau'}^{g},
\label{eq:basic_ribbon}
\end{equation}
where $L_{\tau}^{h} := L^{g}(e_{\tau})$ if the edge $e_{\tau}$ contained in the dual triangle $\tau$ points away from the vertex $v$, $L_{\tau}^{h} := R^{g}(e_{\tau})$ if $e_{\tau}$ points towards the vertex $v$, $T_{\tau'}^{g} := T_{\pm}^{g}(e_{\tau'})$ if the long edge $e_{\tau'}$ of the direct triangle $\tau'$ points in clockwise/counterclockwise
direction with respect to $p$.

\begin{figure}
	\centering
	\includegraphics[width=0.5\textwidth]{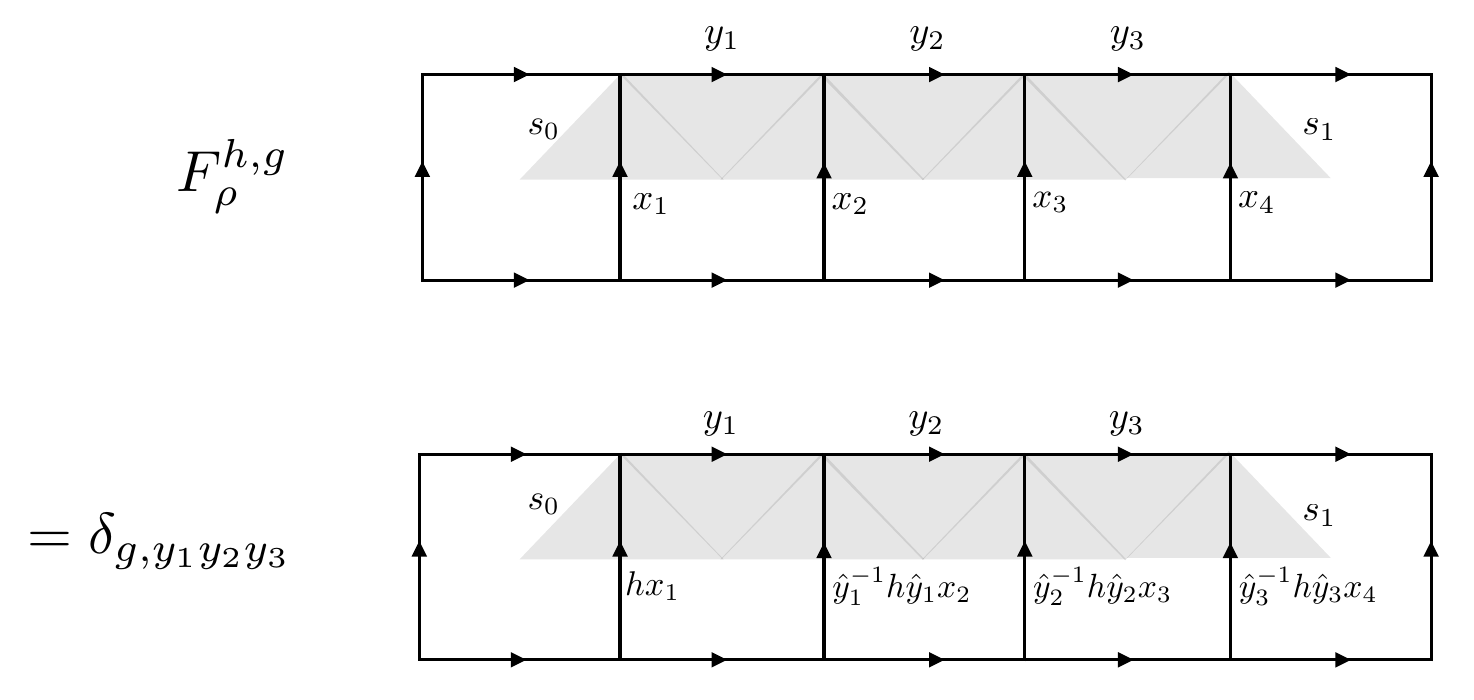}
	\caption{An example of the ribbon operator $F_{\rho}^{h,g}$ for a ribbon $\rho$ with the starting site $s_{0}$ and the ending site $s_{1}$, given in terms of its action on the computational basis states. Here $\hat{y_{j}} = y_{1}...y_{j}$.}
	\label{fig:longribbon}
\label{fig:ribbon_action}
\end{figure}

The ribbon operator on a generic ribbon $\rho$ is defined recursively by the following gluing formula
\begin{equation}
    F_{\rho}^{h,g} = \sum_{k \in G} F_{\rho_{1}}^{h,k} F_{\rho_{2}}^{k^{-1}hk,k^{-1}g}. 
\label{eq:gluingribbon}
\end{equation}
An example of the ribbon operator $F_{\rho}^{h,g}$ is shown in Fig.~\ref{fig:longribbon}, where the action of the ribbon operator on a computational basis state is illustrated. The ribbon operators $F_{\rho}^{h,g}$ create a pair of anyonic excitations at the endpoints of the ribbon $\rho$. However, these excitations may be a superposition of elementary anyons. The elementary anyons are created by ribbon operators in a new basis labeled by $([g],\pi_{g},\boldsymbol{u},\boldsymbol{v})$, where $\boldsymbol{u}=(i,j)$, $\boldsymbol{v}=(i',j')$ such that $i,i' \in \{1,...,|[g]|\}$ index elements of the conjugacy class $[g]$ and $j,j'$ label matrix entries of the irrep $\pi_{g}$. In this basis, we need to define a set $P([g]) = \{ p_{j} \}_{j=1}^{|[g]|}$ of representatives of $G/C_{g}$ such that $c_{j} = p_{j} g p_{j}^{-1}$, where $c_{j}$ enumerate the elements of $[g]$. Every element $g \in G$ can be written in a unique way as $g = p_{j}n$ for some $j \in \{ 1,...,|[g]| \}$ and $n \in C_{g}$. A ribbon operator in the basis $([g],\pi_{g},\boldsymbol{u},\boldsymbol{v})$ is then given by
\begin{equation}
    F_{\rho}^{([g],\pi_{g});(\boldsymbol{u},\boldsymbol{v})} = \frac{\text{dim}(\pi)}{|C_{g}|} \sum_{k \in C_{g}} (\Gamma_{\pi}^{-1}(k))_{j,j'} F_{\rho}^{(c_{i}^{-1},p_{i}kp_{i'}^{-1})}.
\label{eq:ribbon_general}
\end{equation}
As discussed earlier, the pair $([g],\pi_{g})$ labels the anyon type and encodes global degrees of freedom. These labels cannot be changed by local operators at the ends of a ribbon. In contrast, $(\boldsymbol{u},\boldsymbol{v})$ describes local degrees of freedom within each type of anyon and can be changed by applying some local operators at the endpoints. The quantum double model $D(G)$ realizes the $\mathcal{Z}(G)$ topological order\footnote{More precisely, the quantum double model realizes the $\mathcal{Z}(\text{Rep}(G))$ topological order. Since the Drinfeld center $\mathcal{Z}(\text{Rep}(G))$ of the category $\text{Rep}(G)$ of the representation of $G$ is Morita equivalent to $\mathcal{Z}(\text{Vec}(G))$, where $\text{Vec}(G)$ is the category of the $G$-grade vector space.  We will use a short hand notation $\mathcal{Z}(G)$ to denote the topological order where needed.}. 
To form a logical operator, we need a long ribbon operator connecting the same type of the boundaries without violating any Hamiltonian terms on the code block. Any transformation within the $(\boldsymbol{u},\boldsymbol{v})$ space does not change the logical operator type.

\subsection{$D_{4}$ quantum double model}
\label{sec:D4_model}

\begin{figure}
	\centering
	\includegraphics[width=0.3\textwidth]{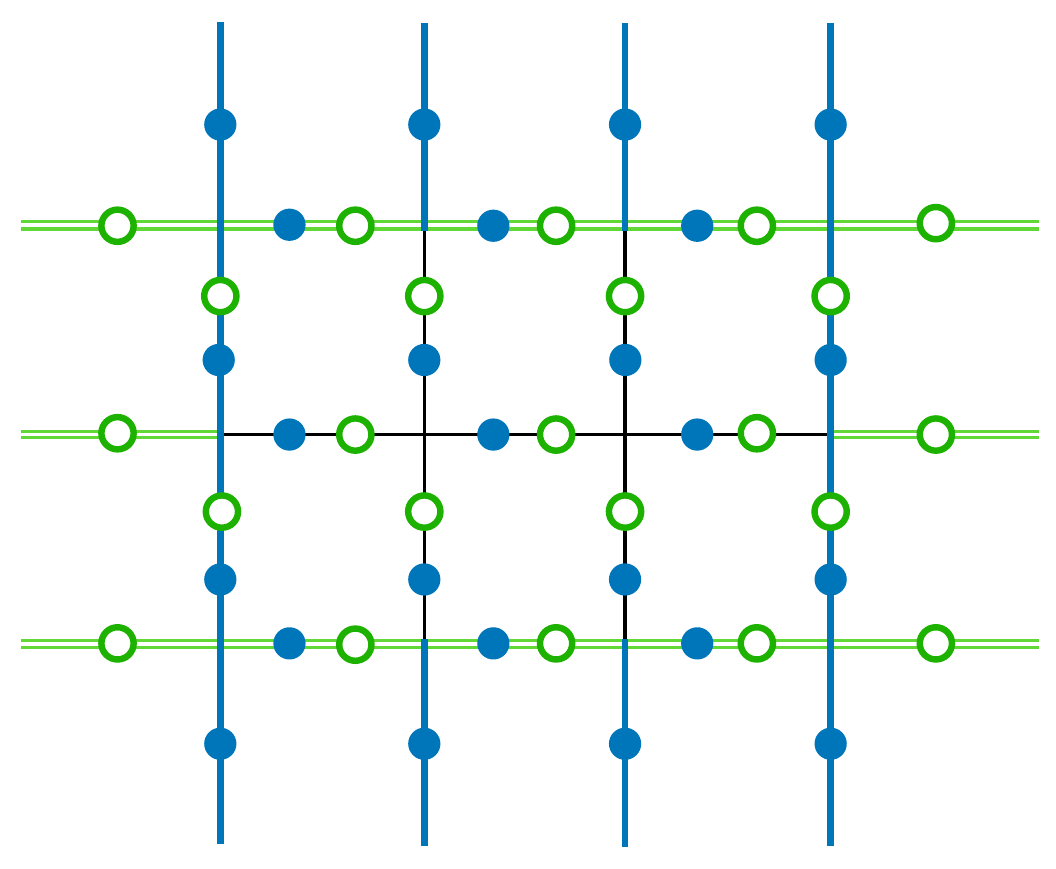}
	\caption{The $D_4$ surface code is defined on a square lattice with a 4-dimensional qudit (filled blue dot) and a qubit (open green dot) at each edge. The boundary edges for the 4-dimensional qudits are indicated by thick blue lines. The boundary edges for the qubits are indicated by double green lines.}
	\label{fig:Lattice_D4}
\end{figure}

Here we provide the details of the $D_{4}$ quantum double model $D(D_{4})$, where $D_{4}$ is the dihedral group of order 8 defined as $< r,s | r^{4}=s^{2}=(rs)^{2} =1>$. The Hilbert space at each edge is $8$-dimensional, and we obtain it from the composition of a qubit and a $4$-dimensional qudit as shown in Fig.~\ref{fig:Lattice_D4}. For the qudit, we denote the shift, clock, and charge conjugation operators as 
\begin{align}
    & \tilde{X} = 
    \begin{pmatrix}
    0 & 0 & 0 & 1 \\
    1 & 0 & 0 & 0 \\
    0 & 1 & 0 & 0 \\
    0 & 0 & 1 & 0 
    \end{pmatrix}, \nonumber\\
    & \tilde{Z} = 
    \begin{pmatrix}
    1 & 0 & 0 & 0 \\
    0 & i & 0 & 0 \\
    0 & 0 & -1 & 0 \\
    0 & 0 & 0 & -i 
    \end{pmatrix}, \nonumber\\
    & \tilde{C} = 
    \begin{pmatrix}
    1 & 0 & 0 & 0 \\
    0 & 0 & 0 & 1 \\
    0 & 0 & 1 & 0 \\
    0 & 1 & 0 & 0 
    \end{pmatrix},
\end{align}
respectively. The former two are the generalized Pauli operators, and the charge conjugation operator satisfies
\begin{equation}
    \tilde{C}\tilde{X}\tilde{C} = \tilde{X}^{-1}, \quad \tilde{C}\tilde{Z}\tilde{C} = \tilde{Z}^{-1}.
\end{equation}
In the regular representation, the left and right multiplications of $D_{4}$ are $8$ by $8$ matrices, which can be explicitly written in terms of the Hilbert spaces of the qubit and the qudit as
\begin{equation}
    L^{r} = I \otimes \tilde{X}, \quad L^{s} = X \otimes \tilde{C},
\end{equation}
\begin{equation}
    R^{r} = \tilde{X}^{-1} \oplus \tilde{X} = \tilde{X}^{-Z}, \quad R^{s} = X \otimes \tilde{I}.
\end{equation}
The vertex terms $A_{v}^{(g)}$ are generated by
\begin{equation}
    A^{(r)}_{v} = \includegraphics[width=.24\linewidth,valign=c]{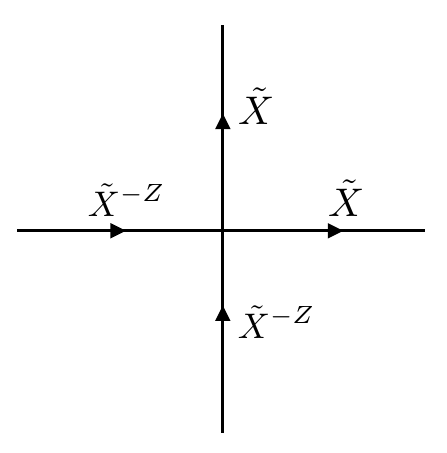}, \quad A^{(s)}_{v} = \includegraphics[width=.22\linewidth,valign=c]{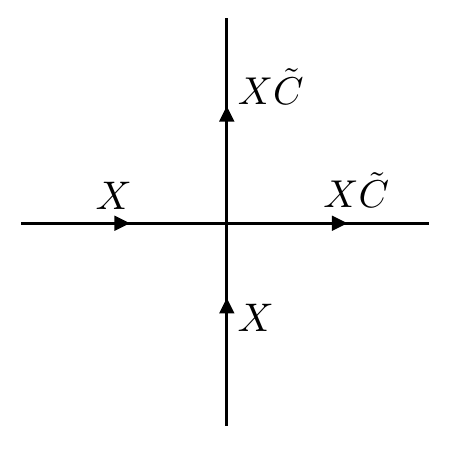},
\end{equation}
and the plaquette terms take the form
$B_{p} = \left( \frac{1+B_{p}^{(r)}+(B_{p}^{(r)})^{2}+(B_{p}^{(r)})^{3}}{4} \right) \cdot \left( \frac{1+B_{p}^{(s)}}{2} \right)$, where
\begin{equation}
    B^{(r)}_{p} = \includegraphics[width=.25\linewidth,valign=c]{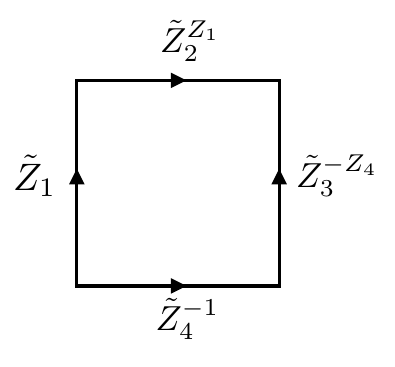}, \quad B^{(s)}_{p} = \includegraphics[width=.21\linewidth,valign=c]{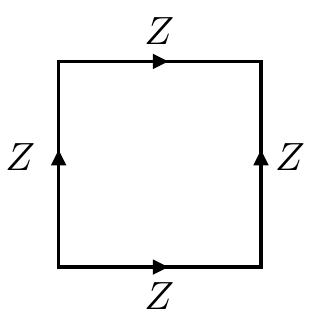}.
\end{equation}

Similar to the standard surface code, the $D_{4}$ quantum double model can be placed on a square lattice with open boundary conditions, which we will refer to as the $D_{4}$ surface code. We choose the following boundary conditions: The $A_{v}^{(r)}$ and $B_{p}^{(r)}$ terms have rough boundary condition on the top and bottom boundaries, and smooth boundary condition on the left and right boundaries; the $A_{v}^{(s)}$ and $B_{p}^{(s)}$ terms have smooth boundary condition on the top and bottom boundaries, and rough boundary condition on the left and right boundaries. The boundary conditions are illustrated in Fig.~\ref{fig:Lattice_D4}. 

We present the conjugacy classes of $D_{4}$ and their centralizers in Table~\ref{tab:conjcent}. 
For the centralizer $C_{r}$, which is isomorphic to $\z_{4}$, we label its 1-dimensional irreducible representations by $\omega_{l} = q^{l}$, where $l=0,1,2,3$ and $q=\exp(i 2 \pi/4)$. For the other centralizers, we take the notation for the irreducible representations of the group to which the centralizer is isomorphic, listed in Tables~\ref{tb:D4} and \ref{tb:D2}. 
Labeled by the conjugacy classes and the irreducible representations of the centralizers, there are 22 anyons in total in the $D_{4}$ quantum double model. These are the same as the anyons in the twisted $\z_{2}^{3}$ theory, since the two models realize the same topological order. We list the quantum dimensions, the topological spins, and the correspondence between the anyons labels in the $D_{4}$ quantum double and the twisted $\z_{2}^{3}$ theory in Table~\ref{tb:D4anyons}. 

\begin{table*}
    \centering
    \begin{tabular}{l l}
    \hline
    Conjugacy class &  Centralizer \\
    \hline
    $[1]=\{1\}$ & $C_{1}=D_4$ \\
    $[r^{2}]=\{r^{2}\}$ & $C_{r^{2}}=D_4$ \\
    $[r] = \{r,r^{3}\}$ & $C_{r}=\z_{4}^{r}$ \\
    $[s] = \{s,r^{2}s\}$ & $C_{s}= D_{2} =\z_{2}^{s} \times \z_{2}^{r^{2}} $ \\
    $[rs] = \{rs,r^{3}s\}$ & $C_{rs}= D_{2} =\z_{2}^{rs} \times \z_{2}^{r^{2}} $\\
    \hline
\end{tabular}
    \caption{Conjugacy classes of the dihedral group $D_{4}$ and their centralizers.}
    \label{tab:conjcent}
\end{table*}

\begin{table*}
\begin{center}
 \begin{tabular}{ c | c c c c c } 
 $D_{4}$ & $[1]$ & $[r^{2}]$ & $[r]$ & $[s]$ & $[rs]$ \\
 \hline
 $J_{0}$ & $1$ & $1$ & $1$ & $1$ & $1$
 \\
 
 $J_{1}$ & $1$ & $1$ & $1$ & $-1$ & $-1$
 \\
 $J_{2}$ & $1$ & $1$ & $-1$ & $1$ & $-1$
 \\
 $J_{3}$ & $1$ & $1$ & $-1$ & $-1$ & $1$
 \\
 $\alpha$ & $2$ & $-2$ & $0$ & $0$ & $0$
 \end{tabular}
\end{center}
\caption{The character table of $D_{4}$.}
\label{tb:D4}
\end{table*}

\begin{table*}
\begin{center}
 \begin{tabular}{ c | c c c c } 
 $D_{2}$ & $[(0,0)]$ & $[(1,0)]$ & $[(0,1)]$ & $[(1,1)]$ \\
 \hline
 $A_{0}$ & $1$ & $1$ & $1$ & $1$
 \\
 
 $A_{1}$ & $1$ & $1$ & $-1$ & $-1$
 \\
 $A_{2}$ & $1$ & $-1$ & $1$ & $-1$
 \\
 $A_{3}$ & $1$ & $-1$ & $-1$ & $1$
 \end{tabular}
\end{center}
\caption{The character table of $D_{2}$.}
\label{tb:D2}
\end{table*}

\begin{table*}
\begin{center}
 \begin{tabular}{ c | c | c | c c } 

$([g],\pi_{g})$ & Anyon label in twisted $\z_{2}^{3}$ gauge theory & Dim. & $\theta$ \\
\hline
$(1,J_{0})$ & $1$ & $1$ & $1$ \\
$(1,J_{1})$ & $e_{RG}$ & $1$ & $1$ \\
$(1,J_{2})$ & $e_{R}$ & $1$ & $1$ \\ 
$(1,J_{3})$ & $e_{G}$ & $1$ & $1$ \\
$(1,\alpha)$ & $m_{B}$ & $2$ & $1$ \\
\hline
$(r^{2},J_{0})$ & $e_{RGB}$ & $1$ & $1$ \\
$(r^{2},J_{1})$ & $e_{B}$ & $1$ & $1$ \\
$(r^{2},J_{2})$ & $e_{GB}$ & $1$ & $1$ \\ 
$(r^{2},J_{3})$ & $e_{RB}$ & $1$ & $1$ \\
$(r^{2},\alpha)$ & $f_{B}$ & $2$ & $-1$ \\
\hline
$(r,\omega_{0})$ & $m_{RG}$ & $2$ & $1$ \\
$(r,\omega_{1})$ & $s_{RGB}$ & $2$ & $i$ \\
$(r,\omega_{2})$ & $f_{RG}$ & $2$ & $-1$ \\ 
$(r,\omega_{3})$ & $\bar{s}_{RGB}$ & $2$ & $-i$ \\
\hline
$(s,A_{0})$ & $m_{GB}$ & $2$ & $1$ \\
$(s,A_{1})$ & $f_{G}$ & $2$ & $-1$ \\
$(s,A_{2})$ & $m_{G}$ & $2$ & $1$ \\ 
$(s,A_{3})$ & $f_{GB}$ & $2$ & $-1$ \\
\hline
$(rs,A_{0})$ & $m_{RB}$ & $2$ & $1$ \\
$(rs,A_{1})$ & $f_{R}$ & $2$ & $-1$ \\
$(rs,A_{2})$ & $m_{R}$ & $2$ & $1$ \\ 
$(rs,A_{3})$ & $f_{RB}$ & $2$ & $-1$ \\

 \end{tabular}
\end{center}
\caption{Anyons in the $D_{4}$ quantum double model can be labeled by a conjugacy class and an irreducible representation of the corresponding centralizer. Anyons in the $D_{4}$ quantum double are equivalent to the anyons in the twisted $\z_{2}^{3}$ gauge theory, listed in the second column. The quantum dimension and the topological spin of each anyon are listed in columns 3 and 4, respectively.}
\label{tb:D4anyons}
\end{table*}

\section{Generating the magic state: continuum analysis}
\label{sec:magic_cat}

The discussion in this section will be based on the algebraic theory of anyons (more precisely, unitary modular tensor categories), which provides a continuum description of the bulk topological order. We treat a code block as a topological order on a patch with a pair of topological gapped boundary conditions. Physically, a topological gapped boundary condition is specified by a set of anyons that can condense on the boundary. A crucial observation is that each condensable anyon string connecting the opposite boundaries corresponds to a logical operator within the code subspace. Therefore, we can describe operations on a logical state during code transformation through manipulations of anyons. Formally, these anyons are described by a Lagrangian algebra $\mathcal{A}$, which is a formal sum of anyons in a bulk topological order $\mathcal{T}$: $\mathcal{A} = \bigoplus_{\alpha \in \mathcal{T}} n_{\alpha} \alpha$, where $n_{\alpha}$ are some non-negative integers. The value $n_\alpha=0$ corresponds to the case where the anyon $\alpha$ cannot condense at the boundary. A positive value means that $\alpha$ can condense at the boundary where $n_\alpha$ specifies the number of inequivalent condensation channels. The collection of all anyons $\{\alpha\}$ with non-vanishing $n_{\alpha}$ is called the maximal set of condensable anyons. We will use this terminology in the following discussion. A brief review of the Lagrangian algebra is given in Appendix~\ref{app:anyon_condensation}. When a topological order $\mathcal{Z}(G)$ is placed on a patch with some topological gapped boundary conditions, the degenerate ground state subspace will be the code subspace. This provides a continuum description of the surface code that realizes the $\mathcal{Z}(G)$ topological order on the patch. We will therefore call such an abstract code in the continuum a $\mathcal{Z}(G)$ surface code. Note that the $G$ surface code in the context of lattice models specifically refers to the Kitaev's quantum double model which realizes the $\mathcal{Z}(G)$ topological order.

We now sketch our main idea of generating the magic state in a $\mathcal{Z}(\z_{2})$ surface code with the help of the $\mathcal{Z}(D_{4})$ surface code. This is based on the observation that a generalized Clifford gate on a $4$-dimensional qudit can introduce a phase on some of the computational states, with the same value as the phase on a qubit computational state produced by a $T$ gate. More explicitly, let us label the logical computational basis in the $\mathcal{Z}(\z_{4})$ surface code by $\{\ket{j}\}, \, j=0,1,2,3$ and consider applying the Fourier transform of the logical $\bar{S}$-gate on the logical state $\ket{0}$:
\begin{align}
    \ket{S_{X}} &:= \bar{H} \cdot \bar{S} \cdot \bar{H}^{\dagger} \ket{0} \nonumber\\
    &= \ket{\omega_{0}} + e^{i \pi/4} \ket{\omega_{1}} - \ket{\omega_{2}} + e^{i \pi/4} \ket{\omega_{3}},
\end{align}
where $\ket{\omega_{j}}$ denotes an eigenstate of the logical $\bar{X}$ operator with the eigenvalue $e^{-i 2\pi j/4}$. $\bar{H}$ and $\bar{S}$ are the logical versions of the gates defined in Eq.~\ref{eq:z4_Fourier} and Eq.~\ref{eq:z4_S}, respectively. We would like to find a series of operations $\mathcal{F}$, potentially with auxiliary degrees of freedom, such that
\begin{equation}
    \mathcal{F}[\ket{S_{X}}] = \ket{T_{X}} = \ket{+} + e^{i\pi/4} \ket{-},
\label{eq:logicaltrans}
\end{equation}
where $\ket{+}$ and $\ket{-}$ are two logical basis states in a surface code that realizes the $\mathcal{Z}(\z_{2})$ topological order and $\ket{T_{X}}$ is the magic state. In the following, we will describe what $\mathcal{F}$ is in the language of anyons.

We consider an abstract $\mathcal{Z}(G)$ surface code with a pair of gapped boundary conditions given by the Lagrangian algebra $\mathcal{A}=\bigoplus_{\alpha \in \mathcal{T}} n_{\alpha} \alpha$ and $\mathcal{B}=\bigoplus_{\beta \in \mathcal{T}} n_{\beta} \beta$. The set of anyon strings $\{\alpha\}$ that condense on the opposite boundaries equipped with the Lagrangian algebra $\mathcal{A}$ forms a set of logical operators $\{L_{\alpha,(i,j)}\}$, where $i,j = 1,2,...,n_{\alpha}$ index the condensation channels. We can then choose to label logical states $\ket{\alpha,(i,j)} = L_{\alpha,(i,j)} \ket{\boldsymbol{1}}$ by this set of anyon strings~\cite{Cong2017}. Here $\boldsymbol{1}$ denotes the trivial anyon and $\ket{\boldsymbol{1}}$ denotes the vacuum state. The set of anyons $\{\beta\}$ that condense on the boundaries equipped with the Lagrangian algebra $\mathcal{B}$ labels another set of logical operators $L_{\beta,(k,l)}$. Their actions on the set of logical states $\{\ket{\alpha,(i,j)}\}$ are given by the mutual braiding: 
\begin{equation}
    L_{\beta,(k,l)} \ket{\alpha,(i,j)} = \frac{S_{\alpha \beta}}{S_{\boldsymbol{1} \alpha}} \ket{\alpha,(i,j)},
\end{equation}
where $S_{\alpha \beta}$ is the $S$ matrix of $\mathcal{Z}(G)$.

As an example, consider the abstract $\mathcal{Z}(\z_{n})$ surface code with the top and bottom boundaries being the e-condensed boundary $\mathcal{A}_{e} = \bigoplus_{j=0}^{n-1} e^{j}$, and the left and the right boundaries being the m-condensed boundary $\mathcal{A}_{m} = \bigoplus_{j=0}^{n-1} m^{j}$. We can choose to denote the aforementioned logical state $\{\ket{\omega_{j}}\}$ equivalently as $\{\ket{e^{j}}\}$. The logical state $\ket{e}$ is an eigenstate of the logical operator $L_{m}$ with eigenvalue $-i$. 
The logical state $\ket{S_{X}}$ and $\ket{T_{X}}$ with the anyon labeling are written as
\begin{equation}
    \ket{S_{X}} = \ket{\boldsymbol{1}} + e^{i \pi/4} \ket{e} - \ket{e^{2}} + e^{i \pi/4} \ket{e^{3}},
\end{equation}
and 
\begin{equation}
    \ket{T_{X}} = \ket{\boldsymbol{1}} + e^{i \pi/4} \ket{e},
\end{equation}
respectively. 

A direct route from a $\mathcal{Z}(\z_{4})$ surface code to a $\mathcal{Z}(\z_{2})$ code is by performing anyon condensation. In this case, we can either condense $e^{2}$ or $m^{2}$ particles to obtain the $\mathcal{Z}(\z_{2})$ surface code. However, we show in Appendix~\ref{app:z4z2} that condensing neither of the two anyons can transform the logical state $\ket{S_{X}}$ into the desired magic state $\ket{T_{X}}$.

\begin{figure*}
	\centering
	\includegraphics[width=0.9\textwidth]{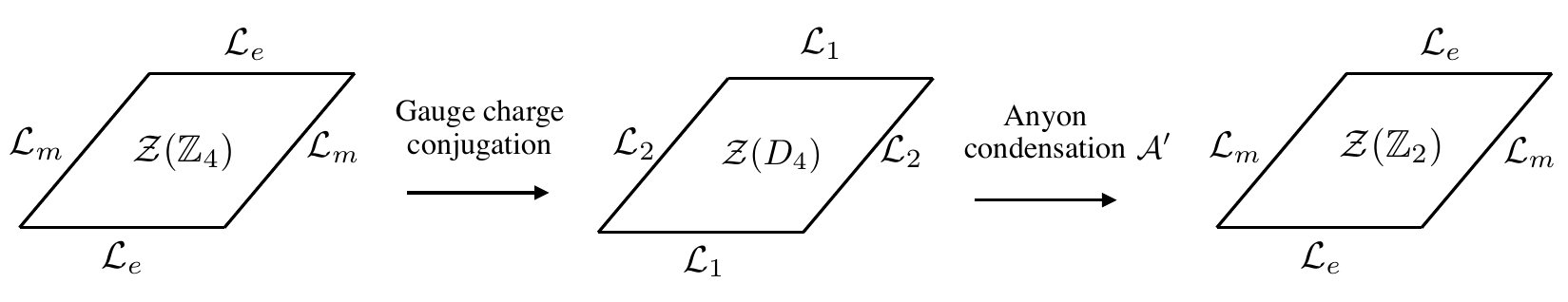}
	\caption{The sequence of topological manipulations transforming the logical state $\ket{S_{X}}$ in a $\mathcal{Z}(\z_{4})$ surface code into the magic state $\ket{T_{X}}$ in a $\mathcal{Z}(\z_{2})$ surface code. The boundary conditions for each surface code are specified by the Lagrangian algebra $\mathcal{L}_{e} = \oplus_{k=0}^{N-1} e^{k}$ and $\mathcal{L}_{m} = \oplus_{k=0}^{N-1} m^{k}$ for $\mathcal{Z}(\z_{N})$ surface code, and $\mathcal{L}_{1} = 1 \oplus e_{R} \oplus m_{B} \oplus m_{G} \oplus m_{GB}$ and $\mathcal{L}_{2} = 1 \oplus e_{B} \oplus e_{RG} \oplus e_{RGB} \oplus 2 m_{RG}$ for $D_{4}$ surface code.}
	\label{fig:code_switch}
\end{figure*}

\begin{figure}
	\centering
	\includegraphics[width=0.4\textwidth]{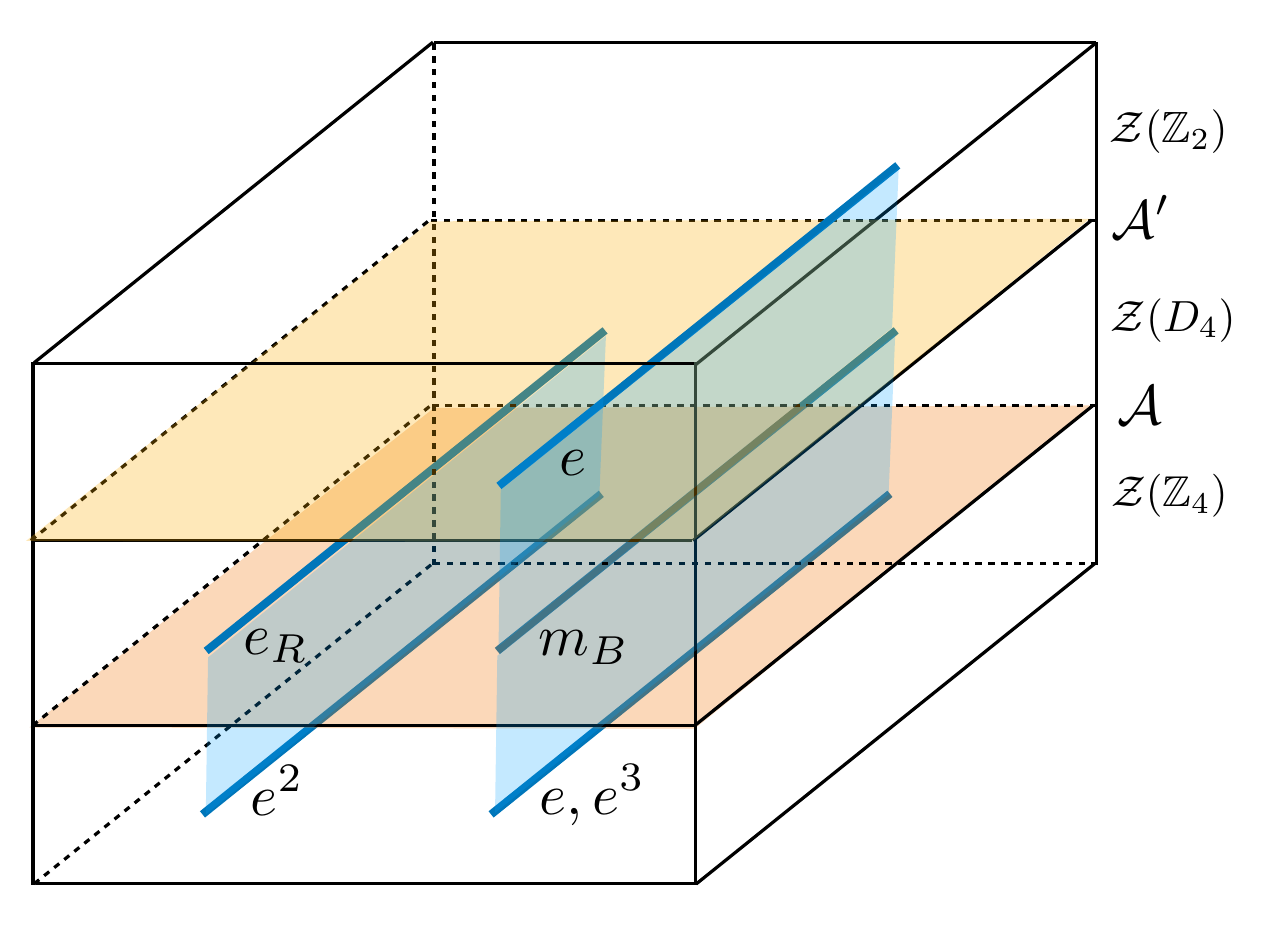}
	\caption{The sequence of topological manipulations seen as a `sandwich' construction. The gapped interface between $\mathcal{Z}(D_{4})$ and $\mathcal{Z}(\z_{4})$ is described by the condensable algebra $\mathcal{A} = 1 \oplus e_{RG}$, and the interface between $\mathcal{Z}(D_{4})$ and $\mathcal{Z}(\z_{2})$ is described by $\mathcal{A}' = 1 \oplus e_{G} \oplus m_{R}$.}
	\label{fig:sandwich}
\end{figure}

Fortunately, this can be remedied by going through a non-Abelian topological order, a $\mathcal{Z}(D_{4})$ surface code, in the middle. We illustrate this procedure in Fig.~\ref{fig:code_switch}. We first gauge the charge conjugation symmetry to go from the $\mathcal{Z}(\z_{4})$ surface code to a $\mathcal{Z}(D_{4})$ surface code. Then the anyons specified by the Lagrangian algebra $\mathcal{A}' = 1 \oplus e_{G} \oplus m_{R}$ are condensed to obtain the $\mathcal{Z}(\z_{2})$ surface code. Formally, the sequence of topological manipulations, which on lattice models describe switching between topological codes, can be viewed as a `sandwich' construction in the spacetime picture as shown in Fig.~\ref{fig:sandwich}. The gapped interfaces between the topological orders are described by a condensable algebra without the Lagrangian condition (see Appendix~\ref{app:anyon_condensation}). Physically, it means that a non-maximal condensable set of anyons are condensed on the interface. The gapped interfaces have been studied extensively for various topological orders~\cite{Bhardwaj2023Club,Bhardwaj2024Hasse}. All the condensable algebra of the $\mathcal{Z}(D_{4})$ topological order are listed in Ref.~\cite{Bhardwaj2024Hasse}. Since the twisted $\z_{2}^{3}$ theory realizes the same topological order as $\mathcal{Z}(D_{4})$, we use the notation in the twisted $\z_{2}^{3}$ theory for anyons in $\mathcal{Z}(D_{4})$ (see Table~\ref{tb:D4anyons}), in order to be consistent with literature. Our construction corresponds to choosing the condensable algebra 
\begin{equation}
    \mathcal{A} = 1 \oplus e_{RG}
\end{equation}
for the gapped interface $\mathcal{I}_{\mathcal{A}}$ between topological orders $\mathcal{Z}(D_{4})$ and $\mathcal{Z}(\z_{4})$, and 
\begin{equation}
    \mathcal{A}' = 1 \oplus e_{G} \oplus m_{R},
\end{equation}
for the interface $\mathcal{I}_{\mathcal{A}'}$ between $\mathcal{Z}(D_{4})$ and $\mathcal{Z}(\z_{2})$ \footnote{These two algebra are labeled as $\mathcal{A}_{1}$ and $\mathcal{A}_{13}$ in Ref.~\cite{Bhardwaj2024Hasse}, respectively.}. In order for the logical information to pass through in the desired way, we choose the Lagrangian algebra for the top and the bottom boundaries of the $\mathcal{Z}(D_{4})$ surface code to be 
\begin{equation}
    \mathcal{L}_{1} = 1 \oplus e_{R} \oplus m_{B} \oplus m_{G} \oplus m_{GB},
    \label{eq:D4_L1}
\end{equation}
and the Lagrangian algebra for the left and the right boundaries to be
\begin{equation}
    \mathcal{L}_{2} = 1 \oplus e_{B} \oplus e_{RG} \oplus e_{RGB} \oplus 2 m_{RG},
    \label{eq:D4_L2}
\end{equation}
as shown in Fig.~\ref{fig:code_switch}. 
Recall that the logical operators correspond to anyon strings condensing on opposite boundaries. We thus have the following logical operators 
\begin{align}
    & \{ L_{e_{R}}, \quad L_{m_{B}}, \quad L_{m_{G}}, \quad L_{m_{GB}}, \nonumber\\
    & \quad L_{e_{B}}, \quad L_{e_{RG}}, \quad L_{e_{RGB}}, \quad L_{m_{RG},(p,q)} \},
\label{eq:L_D4}
\end{align}
where $p, q \in \{1,2\}$ index the two condensation channels on the $\mathcal{L}_{2}$ boundary. 

We will now show that the logical state $\ket{S_{X}}$ will be transformed into the magic state $\ket{T_{X}}$ by the topological manipulations in Fig.~\ref{fig:code_switch}. To go from $\mathcal{Z}(\z_{4})$ to $\mathcal{Z}(D_{4})$, we gauge the charge conjugation symmetry in the $\mathcal{Z}(\z_{4})$, which is the reverse process of condensing $\mathcal{A} = 1 \oplus e_{RG}$ in $\mathcal{Z}(D_{4})$. To see what the logical states $\{\ket{e^{k}}\}$ in the $\mathcal{Z}(\z_{4})$ surface code become in the $\mathcal{Z}(D_{4})$ surface code, we need to know how the corresponding logical operators $\{L_{e^{k}}\}$ map into $\mathcal{Z}(D_{4})$. 
Since the logical operators are anyon strings, we can infer the mapping through topological manipulations from the lift or restriction map between topological orders. 
The lift $\mathcal{Z}(\z_{4}) \rightarrow \mathcal{Z}(D_{4})$ is calculated in Ref.~\cite{Bhardwaj2024Hasse} and summarized in Appendix~\ref{app:anyon_condensation}. 

Specifically from the lift, we will need (a complete version is in Eq.~\eqref{eq:z4liftd4})
\begin{equation}
    e^{2} \rightarrow e_{R} \oplus e_{G}, \quad e \rightarrow m_{B}, \quad e^{3} \rightarrow m_{B}.
\label{eq:liftz4_d4_partial}
\end{equation}
First, since $e_{G}$ can not condense on the boundaries $\mathcal{L}_{1}$ and $\mathcal{L}_{2}$, there is no corresponding logical operator labeled by $e_{G}$. $L_{e^2}$ therefore transforms as \footnote{It is also possible to choose $\mathcal{L}_{1}$ such that there is a logical operator labeled by $e_{G}$. The lift in Eq.~\eqref{eq:liftz4_d4_partial} then does not uniquely determine the mapping of the of logical operators. However, it is still possible to design a gauging procedure such that Eq.~\eqref{eq:e2toeb} holds on the lattice.}
\begin{equation}
    L_{e^{2}} \rightarrow L_{e_{R}}.
\label{eq:e2toeb}
\end{equation}
Second, both $e$ and $e^{3}$ are mapped to the $m_{B}$ anyon. This implies the logical operator $\alpha L_{e} + \beta L_{e^{3}}$ will first be symmetrized and then be mapped to $L_{m_{B}}$ after gauging. In particular, we have 
\begin{equation}
    L_{e} + L_{e^{3}} \rightarrow L_{m_{B}}.
\label{eq:e1e3tomb}
\end{equation}
For the logical state $\ket{S_{X}} = (1+e^{i \pi/4}L_{e}-L_{e^{2}}+e^{i \pi/4}L_{e^{3}})\ket{\boldsymbol{1}}$, applying Eq.~\eqref{eq:e2toeb} and Eq.~\eqref{eq:e1e3tomb}, we have 
\begin{equation}
    \ket{S_{X}} \rightarrow \ket{\boldsymbol{1}} + e^{i \pi/4} \ket{m_{B}} - \ket{e_{R}},
\end{equation}
where $\ket{m_{B}}$ and $\ket{e_{R}}$ are the logical states obtained by applying to the vacuum state the $m_{B}$ and $e_{R}$ string operators that connect the top and bottom boundaries, respectively.

\paragraph{Direct anyon condensation}

To go from $\mathcal{Z}(D_{4})$ to $\mathcal{Z}(\z_{2})$, we can condense the anyons corresponding to the condensable algebra
\begin{equation}
    \mathcal{A}' = 1 \oplus e_{G} \oplus m_{R}.
\label{eq:condAp}
\end{equation}
The restriction map for the anyons associated with the logical operators in Eq.~\eqref{eq:L_D4} is
\begin{equation}
    m_{B} \rightarrow e,
\end{equation}
and the other anyons $e_{R}$, $m_{G}$, and $m_{GB}$ in the Lagrangian algebra associated with the top and bottom boundaries are confined. The restriction map can be obtained from the lift given in Appendix~\ref{app:anyon_condensation}. This leads to the map between the logical operators
\begin{equation}
    L_{m_{B}} \rightarrow L_{e},
\end{equation}
while $L_{e_{R}}$, $L_{m_{G}}$, and $L_{m_{GB}}$ are projected out of the code space. The logical state is therefore transformed as 
\begin{equation}
    \ket{\boldsymbol{1}} + e^{i \pi/4} \ket{m_{B}} - \ket{e_{R}} \rightarrow \ket{T_{X}} = \ket{\boldsymbol{1}} + e^{i \pi/4} \ket{e}.
\end{equation}
In the $\mathcal{Z}(\z_{2})$ surface code, $\ket{\boldsymbol{1}}$ and $\ket{e}$ are orthogonal to each other because they are eigenstates of the logical operators $L_{m}$ with eigenvalues $\pm 1$. We have therefore generated the magic state $\ket{T_{X}}$ in the $\mathcal{Z}(\z_{2})$ surface code from the logical state $\ket{S_{X}}$ in the $\mathcal{Z}(\z_{4})$ surface code through the topological manipulations.   

\begin{figure*}
	\centering
	\includegraphics[width=0.9\textwidth]{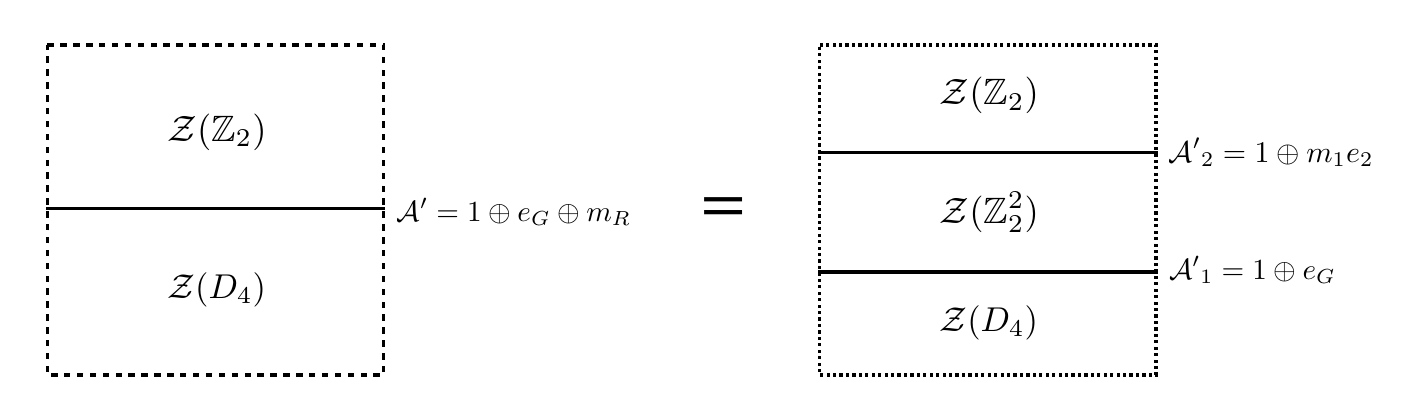}
	\caption{The sequence of condensations $\mathcal{A}_{1}' = 1 \oplus e_{G}$ and $\mathcal{A}_{2}' = 1 \oplus m_{1}e_{2}$ is equivalent to the direct condensation $\mathcal{A}' = 1 \oplus e_{G} \oplus m_{R}$.}
	\label{fig:D4_interfaces}
\end{figure*}

\paragraph{Sequential anyon condensation}
Directly implementing the condensation $\mathcal{A}' = 1 \oplus e_{G} \oplus m_{R}$ on the lattice is complicated since $m_{R}$ is non-Abelian. We can instead perform a sequential condensation equivalent to condensing $\mathcal{A}'$. We first condense the Abelian charge $e_{G}$ with the condensable algebra\footnote{This algebra is labeled as $\mathcal{A}_{5}$ in Ref.~\cite{Bhardwaj2024Hasse}.} 
\begin{equation}
    \mathcal{A}_{1}' = 1 \oplus e_{G},
\label{eq:condAp1}
\end{equation}
which results in the topological order $\mathcal{Z}(\z_{2}^{2})$. This intermediate topological order is the same as two copies of the $\mathcal{Z}(\z_{2})$ topological order in the $\z_{2}$ surface code. The anyons are generated by $\{e_{1},m_{1},e_{2},m_{2}\}$. Then, we condense 
\begin{equation}
    \mathcal{A}_{2}' = 1 \oplus m_{1}e_{2}
\label{eq:condAp2}
\end{equation}
to obtain $\mathcal{Z}(\z_{2})$. The sequence of condensing $\mathcal{A}_{1}'$ and $\mathcal{A}_{2}'$ is equivalent to condensing $\mathcal{A}' = 1 \oplus e_{G} \oplus m_{R}$ as shown in Fig.~\ref{fig:D4_interfaces}. At both stages, the anyons to condense are Abelian and there exist corresponding operations on the lattice.

The relevant restriction map $\mathcal{Z}(D_{4}) \rightarrow \mathcal{Z}(\z_{2}^{2})$ is given by
\begin{equation}
    e_{R} \rightarrow e_{1}, \quad m_{B} \rightarrow e_{2} \oplus e_{1}e_{2},
\end{equation}
which can be derived from the lift provided in Appendix~\ref{app:anyon_condensation}. This gives the map between the logical operators 
\begin{equation}
    L_{e_{R}} \rightarrow L_{e_{1}}, \quad L_{m_{B}} \rightarrow L_{e_{2}}+L_{e_{1}e_{2}},
\end{equation}
while $L_{m_{G}}$ and $L_{m_{GB}}$ are confined and projected out of the code space. The transformation of the logical state is thus
\begin{align}
    & \ket{\boldsymbol{1}} + e^{i \pi/4} \ket{m_{B}} - \ket{e_{R}} \nonumber\\
    & \rightarrow \ket{\boldsymbol{1}} + e^{i \pi/4} \ket{e_{2}} + e^{i \pi/4} \ket{e_{1}e_{2}} - \ket{e_{1}}.
\end{align}
We remark that this logical state encoded in the topological order $\mathcal{Z}(\z_{2}^{2})$ already contains magicness, as it can be regarded as obtained by a non-Clifford gate and some Clifford gates acting on two logical qubits. From this state, we can disentangle the two qubits and extract a magic state using Clifford gates. Alternatively, we can further condense $\mathcal{A}_{2}'=1\oplus m_{1}e_{2}$ in $\mathcal{Z}(\z_{2}^{2})$ to obtain $\mathcal{Z}(\z_{2})$. The anyons $e_{1}$ and $e_{1}e_{2}$ that braid non-trivially with $m_{1}e_{2}$ will be confined after condensation. The relevant restriction map is given by
\begin{equation}
    e_{2} \rightarrow e,
\end{equation}
and the map between the logical operators is
\begin{equation}
    L_{e_{2}} \rightarrow L_{e},
\end{equation}
where $L_{e_{1}}$ and $L_{e_{1}e_{2}}$ are projected out of the code space. The logical state is transformed as 
\begin{align}
   & \ket{\boldsymbol{1}}  + e^{i \pi/4} \ket{e_{2}} + e^{i \pi/4} \ket{e_{1}e_{2}} - \ket{e_{1}} \nonumber\\
   & \rightarrow \ket{T_{X}} = \ket{\boldsymbol{1}} + e^{i \pi/4} \ket{e}.
\end{align} 
The result of the sequence of $\mathcal{A}_{1}'$ and $\mathcal{A}_{2}'$ condensations is equivalent to the result of the $\mathcal{A}' = 1 \oplus e_{G} \oplus m_{R}$ condensation.

\section{Generating the magic state: detailed procedure on the lattice}
\label{sec:magic_lattice}

To achieve the desired transformation on the logical information, the topological manipulations consist of gauging the charge conjugation symmetry in the $\mathcal{Z}(\mathbb{Z}_{4})$ surface code, condensing $e_{G}$ in the $\mathcal{Z}(D_{4})$ surface code, and either applying a transversal entangling gate or subsequently condensing $m_1e_2$, as discussed in Sec.~\ref{sec:magic_cat}. In this section, we will discuss the lattice realization of these topological manipulations.


The lattice model that realizes the $\mathcal{Z}(\z_{4})$ topological order is the standard $\z_{4}$ surface code reviewed in Sec.~\ref{sec:zn_surface_code}. We will discuss how to gauge the charge conjugation symmetry in the $\z_{4}$ surface code and the corresponding transformation on the logical information. The result of gauging the charge conjugation symmetry is the $D_{4}$ surface code reviewed in Sec.~\ref{sec:D4_model}, corresponding to the abstract $\mathcal{Z}(D_{4})$ surface code in the continuum description. 
After performing the anyon condensation $\mathcal{A}_{1}'$ in Eq.~\eqref{eq:condAp1}, we obtain a $\mathcal{Z}(\z_{2}^2)$ topological order, which can be transformed into two $\z_{2}$ surface codes through local basis change. From this, we can either (1) apply a transversal gate to disentangle the qubits encoded in the two $\z_{2}$ surface codes, or (2) further condense the anyon $\mathcal{A}_{2}'$ in Eq.~\eqref{eq:condAp2} to reach the $\z_{2}$ surface code. With Option (1) we can extract a magical state encoded in one of the two $\z_{2}$ surface codes. Option (2) leads to a commuting projector Hamiltonian realizing the same topological order as the standard $\z_{2}$ surface code, which we call the \emph{condensed $\z_{2}$ surface code} and discuss in Sec.~\ref{sec:condensed_z2_surface_code}. This can be turned into the standard $\z_{2}$ surface code via local Clifford gates. A magic state encoded in the standard or condensed $\z_{2}$ surface code can be used for executing a logical $T$ gate in the standard $\z_{2}$ surface code block through teleportation.
Following the notation in Sec.~\ref{sec:review_Kitaev}, we will denote an operator in the qubit Hilbert space, an operator in the 4-dimensional qudit, and a logical operator in the initial $\z_{4}$ surface code or the final $\z_{2}$ surface code by $O$, $\tilde{O}$, and $\bar{O}$ respectively.

\subsection{From $\z_{4}$ surface code to $D_{4}$ surface code by gauging charge conjugation} 
\label{sec:gauge_c}

\begin{figure}
	\centering
	\includegraphics[width=0.3\textwidth]{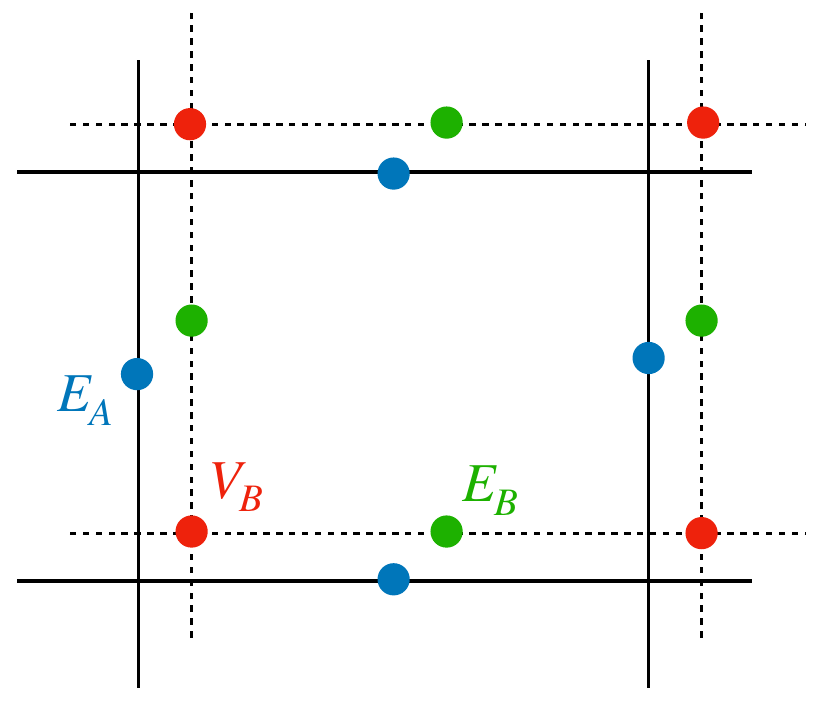}
	\caption{Ancilla qubits (marked by red and green) are prepared in the $\ket{+}$ state and placed at the vertices and edges of a shifted square lattice. The set of edges in the original lattice is denoted by $E_{A}$, the set of vertices and the set of edges in the shifted square lattice are denoted by $V_{B}$ and $E_{B}$, respectively.}
	\label{fig:gauge_c_lattice}
\end{figure}

To gauge the charge conjugation symmetry in the $\z_{4}$ surface code, we follow a similar procedure as in Ref.~\cite{Verresen2021}. We first add ancilla qubits in the $\ket{+}$ state sitting at the vertices and edges of a shifted square lattice as shown in Fig~\ref{fig:gauge_c_lattice}. This lattice is shifted for clear visualization, and will be placed on top of the original lattice in the end to recover a regular square lattice. 
We choose the top and bottom boundaries of the shifted square lattice to be smooth, and the left and right boundaries to be rough, which is opposite to the choice for the original square lattice. Such a boundary choice is important in order to obtain the boundary conditions $\mathcal{L}_{1}$ and $\mathcal{L}_{2}$ in Eq.~\eqref{eq:D4_L1} and Eq.~\eqref{eq:D4_L2} for the $D_{4}$ quantum double model obtained by gauging. We denote the set of edges in the original lattice by $E_{A}$, and the set of vertices and the set edges in the shifted square lattice by $V_{B}$ and $E_{B}$, respectively. 

We first implement a controlled-charge-conjugation operation on each pair of qubits on a vertex in $V_{B}$ and a nearest neighboring edge in $E_{A}$
\begin{equation}
    U_{C\mathcal{\tilde{C}}} = \prod_{\langle v,l \rangle} C\mathcal{\tilde{C}}_{v,l},
\end{equation}
where $\langle v,l \rangle$ denotes nearest neighbor pair of $v \in V_{B}$ and $l \in E_{A}$, and $C\mathcal{\tilde{C}}_{v,l}$ is the controlled-$\mathcal{\tilde{C}}$ gate where $v$ is the control. This unitary introduces for each $v \in V_B$ a new stabilizer 
\begin{equation}
   \includegraphics[width=.25\linewidth,valign=c]{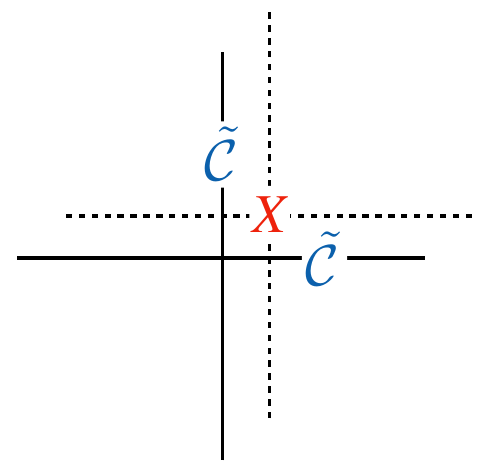}.
\end{equation}
The charge conjugation symmetry $S_{C} = \prod_{l \in E_{A}} \mathcal{\tilde{C}}_{l}$ is equivalent to the the $\z_{2}$ symmetry $S_{\z_{2}} =\prod_{v \in V_{B}} X_{v}$ up to the new stabilizers. We then proceed to gauge the $\z_{2}$ symmetry $S_{\z_{2}}$, which is equivalent to gauging $S_{C}$. 

To gauge the $\z_{2}$ symmetry $S_{\z_{2}}$, we implement the unitary
\begin{equation}
    U_{CZ} =\prod_{\langle v, l \rangle} CZ_{v,l},
\label{eq:ucz}
\end{equation}
where $\langle v,l \rangle$ denotes any nearest neighbor pair of $v \in V_{B}$ and $l \in E_{B}$. If we interpret the qubits in $V_{B}$ as the $\z_{2}$ matter field and the qubits in $E_{B}$ as the $\z_{2}$ gauge field, $U_{CZ}$ is the unitary that couples the $\z_{2}$ matter to the $\z_{2}$ gauge field. The stabilizers at this stage are given by
\begin{align}
   \includegraphics[width=.32\linewidth,valign=c]{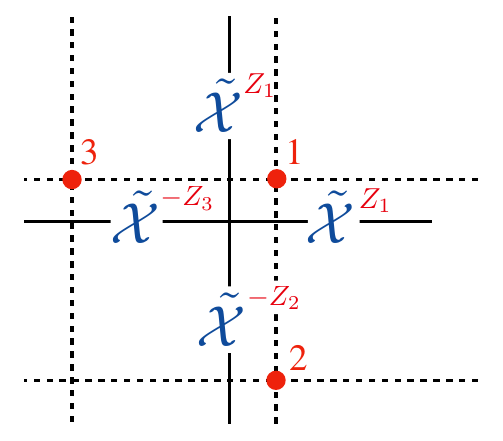}, 
   \quad
   \includegraphics[width=.32\linewidth,valign=c]{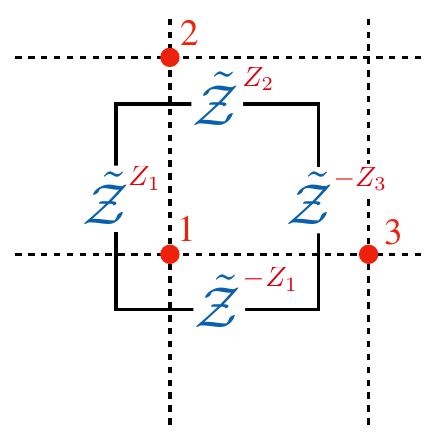},
   \nonumber\\
   \includegraphics[width=.32\linewidth,valign=c]{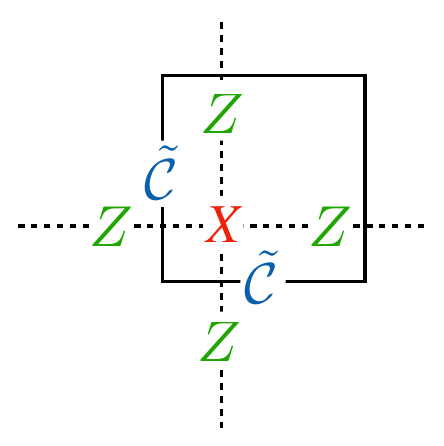},
   \quad
   \includegraphics[width=.25\linewidth,valign=c]{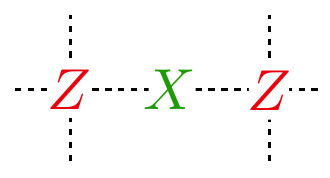},
   \quad
   \includegraphics[width=.12\linewidth,valign=c]{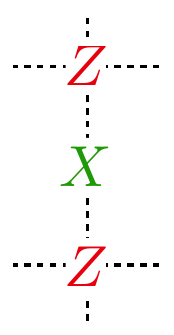}.
\label{eq:stabgauge}
\end{align}
We denote the first order-four stabilizer as $O_{v}$ and the second order-four stabilizer as $O_{p}$ below. 

The next step is to measure the qubit at each vertex in $V_{B}$ in the $X$ basis. The order-four stabilizers $O_{v}$ and $O_{p}$ in Eq.~\eqref{eq:stabgauge} do not commute with the $X$ measurement. However, the projector $\prod_{i=0}^{3}O_{v}^{i}/4$ formed by the stabilizer $O_{v}$ does commute with the measurement. Similarly, the projector $\prod_{i=0}^{3}O_{p}^{i}/4$ commutes with the measurement. We now shift the lattice formed by $E_B$ back to coincide with $E_A$ so that there is one qubit and one $4$-dimensional qudit on each edge of the square lattice. After the $X$ measurement, we apply a Hadamard gate $H$ to each qubit in $E_{B}$, in order to restore the form of the quantum double model. Following the argument in Ref.~\cite{Verresen2021}, the result is a commuting projector model with the following projectors:
\begin{equation}
    A^{1}_{v} = \frac{1}{4} \left( 1+ \includegraphics[width=.2\linewidth,valign=c]{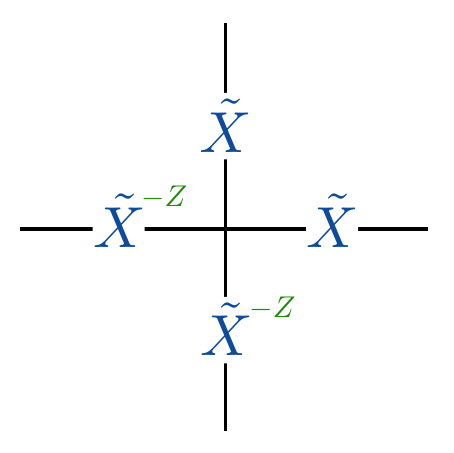} + \includegraphics[width=.2\linewidth,valign=c]{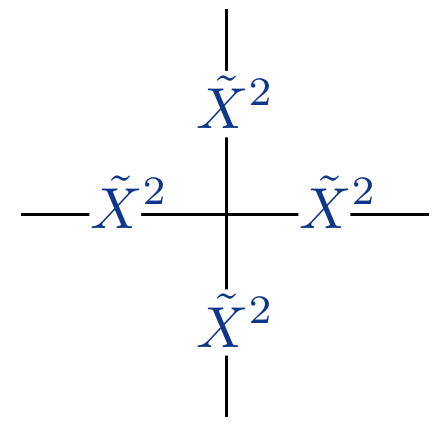} + \includegraphics[width=.2\linewidth,valign=c]{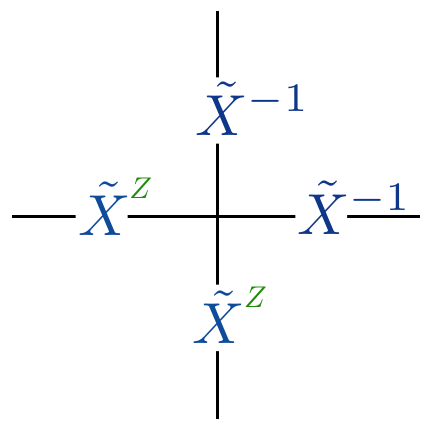} \right),
\end{equation}
\begin{equation}
    A^{2}_{v} = \frac{1}{2} \left( 1+ \includegraphics[width=.2\linewidth,valign=c]{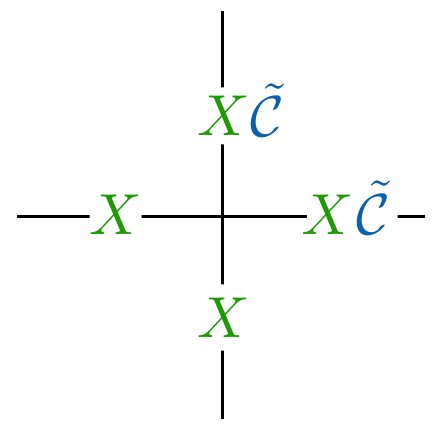}\right),
\end{equation}
\begin{equation}
    B^{1}_{p} = \frac{1}{4} \left( 1+ \includegraphics[width=.16\linewidth,valign=c]{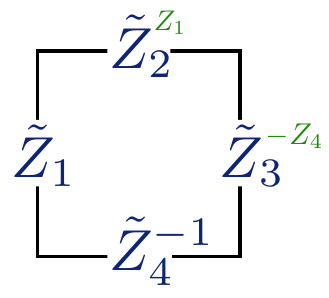} + \includegraphics[width=.16\linewidth,valign=c]{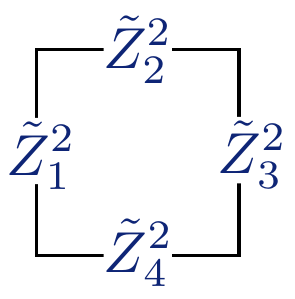} + \includegraphics[width=.16\linewidth,valign=c]{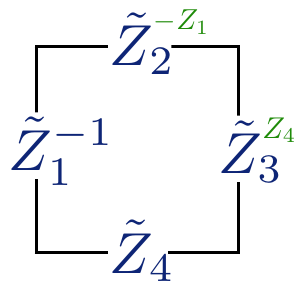} \right),
\end{equation}
\begin{equation}
    B^{2}_{p} = \frac{1}{2} \left( 1+ \includegraphics[width=.13\linewidth,valign=c]{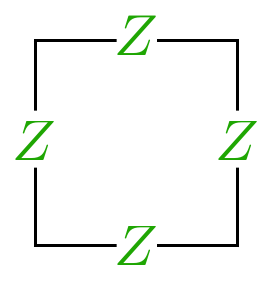}\right).
\end{equation}
We see that these are exactly the projectors of the $D_{4}$ quantum double model in Sec.~\ref{sec:D4_model}. The gauging procedure is an adaptive finite-depth local-unitary circuit which we refer to as the gauging map $\hat{G}$. The gauging map $\hat{G}$ can be written more concisely as
\begin{equation}
    \hat{G} \ket{\Psi} := H_{E_{B}} \text{KW}^{\z_{2}}_{V_B, E_B} \left( U_{C\tilde{\mathcal{C}}} \ket{\Psi} \ket{+}_{V_{B}} \right),
\end{equation}
where 
\begin{equation}
     \text{KW}^{\z_{2}}_{V_B, E_B} (\ket{\cdot}_{V_B}) = \me{+}{_{V_{B}} U_{CZ} \ket{\cdot}_{V_B} } {+}_{E_{B}}
\label{eq:KW}
\end{equation}
is the Kramers-Wannier map for the $\z_{2}$ group~\cite{Verresen2021,Tantivasadakarn2023Long}, and $H_{E_{B}} = \prod_{l \in E_{B}} H_{l}$.

With our choice of boundary condition, the projectors $A_{v}^{r}$ and $B_{p}^{r}$ have rough boundary condition on the top and bottom boundaries, and smooth boundary condition on the left and right boundaries. For the projectors $A_{v}^{m}$ and $B_{p}^{m}$, the top and bottom boundaries are smooth while the left and right boundaries are rough. This means that the top and bottom boundaries condense the anyons in $\mathcal{L}_{1}$, and the left and right boundaries condense the anyons in $\mathcal{L}_{2}$.

\subsubsection{Transformation of logical information}
Now we discuss the transformation of the logical operators from the $\z_{4}$ surface code to the $D_{4}$ surface code. The logical $\bar{Z}^{2}$ operator is given by the $e^{2}$ string operator of the form $\prod_{l \in \rho} \tilde{Z}^{2}_l$, where $\rho$ is a path connecting the rough boundaries. After the gauging procedure, $\prod_{l \in \rho} \tilde{Z}^{2}_l$ stays the same. The ribbon operator of the $e_{R}$ particle is of the same form, as shown in Appendix~\ref{app:D4_model_ribbon} using the general expression of the ribbon operator Eq.~\eqref{eq:ribbon_general}. Therefore, we find that the logical $\bar{Z}^{2}$ operator becomes the logical $L_{e_{R}}$ operator through gauging. 

The logical $\bar{Z}$ and $\bar{Z}^{3}$ operators will combine into the logical $m_{B}$ operator after gauging. To see this, consider a logical $\bar{Z}$ operator of the form $\bar{Z} = \tilde{Z}_{[12]}\tilde{Z}_{[23]}\tilde{Z}_{[34]}\tilde{Z}_{[45]}...$ along a path $\rho$ that starts from the bottom boundary and ends at the top boundary, where $[ij]$ denotes an edge in $E_A$ connecting vertices $i$ and $j$. Applying $U_{C\tilde{\mathcal{C}}}$, we have 
\begin{align}
    U_{C\tilde{\mathcal{C}}} \bar{Z} U_{C\tilde{\mathcal{C}}}^{\dagger} &= \tilde{Z}_{[12]}^{Z_{1}}\tilde{Z}_{[23]}^{Z_{2}}\tilde{Z}_{[34]}^{Z_{3}}\tilde{Z}_{[45]}^{Z_{4}}... \nonumber
    \\
    &=\left( \tilde{Z}_{[12]}\tilde{Z}_{[23]}^{Z_{1}Z_{2}}\tilde{Z}_{[34]}^{Z_{1}Z_{3}}\tilde{Z}_{[45]}^{Z_{1}Z_{4}}... \right)^{Z_{1}},
\end{align}
where we use the same labels for edges in $E_A$ and $E_B$. 

Applying $U_{CZ}$ in Eq.~\eqref{eq:KW} and using the stabilizer condition $Z_{i}X_{[ij]}Z_{j} = 1$ for each edge, we obtain
\begin{equation}
    \mathcal{O}_{\rho} = \left( \tilde{Z}_{[12]}\tilde{Z}_{[23]}^{X_{[12]}}\tilde{Z}_{[34]}^{X_{[12]}X_{[23]}}\tilde{Z}_{[45]}^{X_{[12]}X_{[23]}X_{[34]}}... \right)^{Z_{1}},
\end{equation}
which does not commute with the $X$ measurement in Eq.~\eqref{eq:KW}. However, if we start from $\bar{Z} + \bar{Z}^{3}$ operator, then $\mathcal{O}_{\rho} + \mathcal{O}_{\rho}^{\dagger}$ will survive from the $X$ measurement.
The result of the logical $\bar{Z} + \bar{Z}^{3}$ operator after the gauging transformation is
\begin{align}
    &\left( \tilde{Z}_{[12]}\tilde{Z}_{[23]}^{Z_{[12]}}\tilde{Z}_{[34]}^{Z_{[12]}Z_{[23]}}\tilde{Z}_{[45]}^{Z_{[12]}Z_{[23]}Z_{[34]}}... + \mathrm{h.c.} \right) \nonumber\\
    &= 2\left( F_{\rho}^{1,1} - F_{\rho}^{1,r^{2}} + F_{\rho}^{1,s} - F_{\rho}^{1,r^{2}s} \right), \nonumber
    \\
    &= 2\left( \text{Tr}[F_{\rho}^{m_{B};(\boldsymbol{u},\boldsymbol{v})}] + \text{ATr}[F_{\rho}^{m_{B};(\boldsymbol{u},\boldsymbol{v})}] \right),
\label{eq:mb_tratr}
\end{align}
where the equality in the first line is obtained using Eq.~\eqref{eq:mb_ribbons_proj} in Appendix~\ref{app:D4_model_ribbon}. The trace in the second line is over the local degrees of freedom $(\boldsymbol{u},\boldsymbol{v})$, and $\text{ATr}[M] = \sum_{j=1}^{n} M_{j,n-j+1}$ is the anti-trace of a matrix $M$. This is the ribbon operator for $m_B$ (see Appendix~\ref{app:D4_model_ribbon}) \footnote{We remark that the ribbon operator of $m_{B}$ is in a particular superposition of the internal indices $(\boldsymbol{u},\boldsymbol{v})$, which is related to our choice of the gauging procedure (specifically the choice of measuring $\ket{+}_{V_{B}}$)~\cite{Ren2024}. The anyon type of the ribbon operator is independent of the choice of gauging so this choice does not affect the conclusion on how anyons transform.}.

As a consistency check, we can condense $e_{RG}$ in the $D_{4}$ quantum double and go back to the $\z_{4}$ surface code. This condensation is projecting all the qubits in $E_B$ onto the $\ket{0}$ state. We can be implement it by measuring all the short string operators, the Pauli $Z$ operator on each edge that create a pair of $e_{RG}$ particles, followed by post-processing. One can check that the commuting projectors in the $D_{4}$ quantum double reduce to the stabilizers in the $\z_{4}$ surface code. In particular, the ribbon operator $F_{\rho}^{m_{B}}$ in the projected subspace takes the form $\sim (\prod_{l}\tilde{Z}_{l} + \mathrm{h.c})$, creating a pair of non-simple anyons, each of which is a superposition of $e$ and $e^{3}$ anyons. Therefore, we find that $m_{B}$ particle splits into $e$ and $e^{3}$ particles as expected. 

If we write the logical state $\ket{S_{X}}$ in the $\z_4$ surface code as 
\begin{equation}
    \ket{S_{X}} = \left( 1 + e^{i \pi/4} \bar{Z} - \bar{Z}^{2} + e^{i \pi/4} \bar{Z}^{3} \right) \ket{\omega_{0}},
\end{equation}
the logical state after gauging becomes
\begin{equation}
    \left( 1 + 2 e^{i \pi/4} F_{\rho}^{m_{B}} - F_{\rho}^{e_{R}}  \right) \ket{\Omega},
\end{equation}
where $\ket{\Omega}$ is the ground state of the $D_{4}$ quantum double model obtained from $\ket{\omega_{0}}$ after gauging. The logical state $\ket{S_{X}}$ is mapped into a logical state in the $D_{4}$ quantum double model.

\subsection{From $D_{4}$ surface code to $\z_{2}^{2}$ surface code by anyon condensation}
\label{sec:eg_cond}

To condense the Abelian charge $e_{G}$ in the $D_{4}$ surface code, we measure all the short ribbon operators that create a pair of $e_{G}$ particles and post-processing. 
In Sec.~\ref{app:D4_model_ribbon} we have shown that this corresponds to applying the projector 
\begin{equation}
    P_{l}^{e_{G}} = \frac{1+\tilde{Z}_{l}^{2}Z_{l}}{2}. \label{eq:P_eG}
\end{equation}
on each edge $l$. This projects the local Hilbert space at the edge $l$ to a subspace spanned by
\begin{equation}
    \{ \ket{\boldsymbol{1}}, \ket{r^{2}}, \ket{rs}, \ket{r^{3}s} \}, 
\end{equation}
which are the $4$ states labeled by the subgroup $M = \z_{2}^{rs} \times \z_{2}^{r^{2}}$ of $D_4$. This basis can be identified with a basis of the tensor product form
\begin{equation}
    \{ \ket{\boldsymbol{1}} \rightarrow \ket{00}, \ket{r^{2}} \rightarrow \ket{01}, \ket{rs} \rightarrow \ket{10}, \ket{r^{3}s} \rightarrow \ket{11} \}. 
\label{eq:z22subspace}
\end{equation}
In this subspace, the vertex operators $A_{v}^{m}$ become
\begin{align}
    A_{v}^{rs} =  \includegraphics[width=.2\linewidth,valign=c]{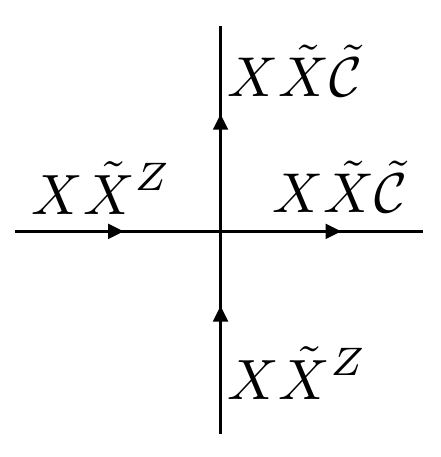} \Rightarrow \includegraphics[width=.2\linewidth,valign=c]{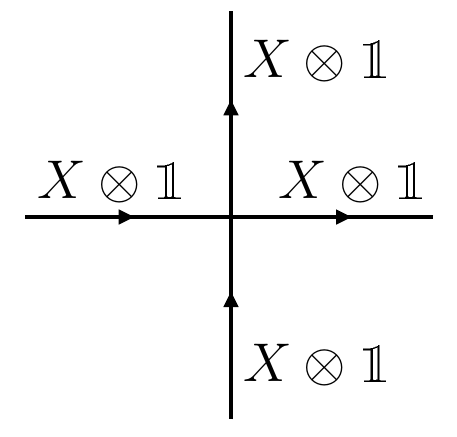}, \nonumber
    \\
    A_{v}^{r^{2}} =  \includegraphics[width=.2\linewidth,valign=c]{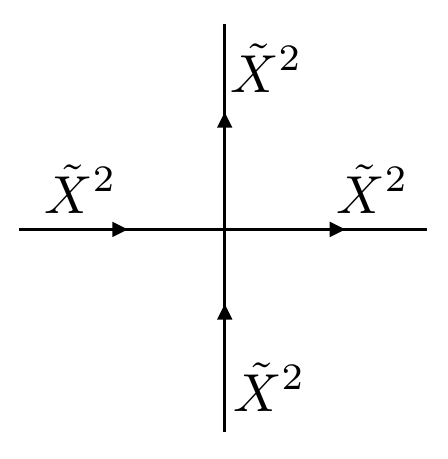} \Rightarrow \includegraphics[width=.2\linewidth,valign=c]{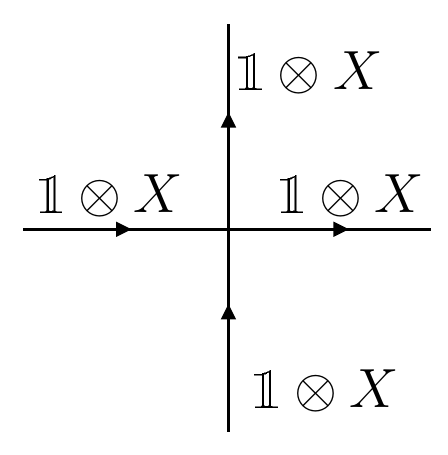}, \nonumber
    \\
    A_{v}^{r^{3}s} =  \includegraphics[width=.2\linewidth,valign=c]{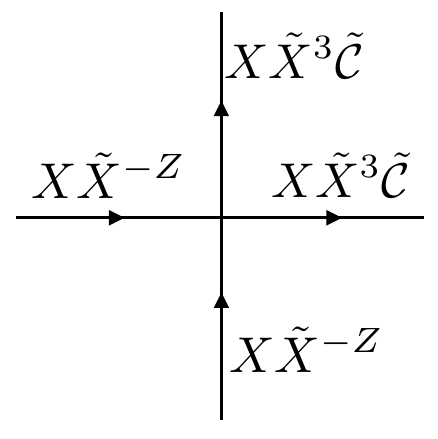} \Rightarrow \includegraphics[width=.2\linewidth,valign=c]{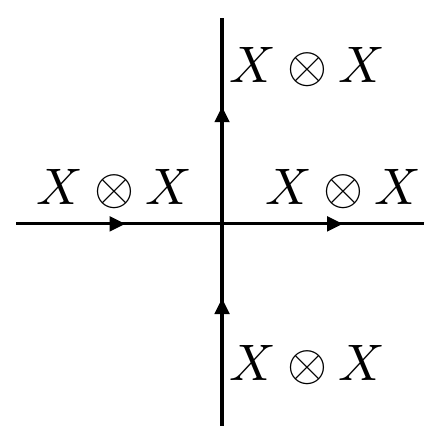},
\end{align}
where the double arrow means projecting to the subspace in Eq.~\eqref{eq:z22subspace}. 
The vertex terms in the Hamiltonian now take the form
\begin{equation}
    A_{v}^{M} = \frac{1}{4} \sum_{m \in M} A_{v}^{m}.
\end{equation}
The $B_{p}$ terms remain unchanged since they commute with the projectors. The ground states are common eigenstates of the projectors 
\begin{equation}
   \{ A_{v}^{M}, B_{p}, P_{l}^{e_{R}} \}.     
\end{equation}
This anyon condensation is an example of the class of condensate in Ref.~\cite{Bombin2008}, where the condensed phase is studied in details. After condensing $e_{G}$, the model realizes the $\mathcal{Z}(\z_{2}^{2})$ topological order. 
We recognize that the vertex operators $A_{v}^{m}$ in the projected subspace take the standard form in the $\z_{2}^{2}$ toric code, which allows us to identify the excitations to the charges in the $\mathcal{Z}(\mathbb{Z}_{2}^{2})$ topological order. Specifically, the excitation that violates $A_{v}^{r^{3}s}$ and $A_{v}^{rs}$ terms will be called $e_{1}$, and the excitation that violates $A_{v}^{r^{2}}$ and $A_{v}^{rs}$ terms will be called $e_{2}$. 

We need to keep track of the logical information after condensation, which amounts to the mapping of the anyon excitations. 
Let us first consider a short ribbon operator on a triangle $\tau$, $F_{\tau}^{e_{R}}$ or $F_{\tau}^{m_{B}}$, that creates a pair of $e_{R}$ or $m_{B}$ particles. As shown in Appendix~\ref{app:D4_model_ribbon}, it is the identity on a dual triangle and non-trivial on a direct triangle. Below we will use only the long edge to label the location of the short ribbon operator. $F_{[12]}^{e_{R}}$ on a direct triangle is $\tilde{Z}_{[12]}^{2}$, which preserves its form after anyon condensation. In the projected subspace of Eq.~\eqref{eq:z22subspace}, it becomes 
\begin{equation}
  F_{[12]}^{e_{R}} = \tilde{Z}_{[12]}^{2} \Rightarrow Z \otimes \mathbbm{1},
\label{eq:se1}
\end{equation}
which violates the vertex operators $A_{v}^{rs}$ and $A_{v}^{r^{3}s}$ at the vertices $v_{1}$ and $v_{2}$. We thus find that $e_{R}$ particle becomes $e_{1}$ particle in the $\mathbb{Z}_{2}^{2}$ surface code. 

On a direct triangle,
\begin{equation}
    F_{[12]}^{m_{B}} = F_{[12]}^{1,1} - F_{[12]}^{1,r^{2}} + F_{[12]}^{1,s} - F_{[12]}^{1,r^{2}s}.
\end{equation}
After the anyon condensation, only the combination $F_{[12]}^{1,1} - F_{[12]}^{1,r^{2}}$ survives, which can be written as 
\begin{align}
    F_{[12]}^{1,1} - F_{[12]}^{1,r^{2}} &= \frac{1}{2} \left( F_{[12]}^{1,1} - F_{[12]}^{1,r^{2}} + F_{[12]}^{1,rs} - F_{[12]}^{1,r^{3}s}  \right) \nonumber\\
    &+ \frac{1}{2} \left( F_{[12]}^{1,1} - F_{[12]}^{1,r^{2}} - F_{[12]}^{1,rs} + F_{[12]}^{1,r^{3}s}  \right).
\end{align}
We focus on the combination of the short string operators in each parentheses in the projected subspace
\begin{align}
     \left( F_{[12]}^{1,1} - F_{[12]}^{1,r^{2}} + F_{[12]}^{1,rs} - F_{[12]}^{1,r^{3}s}  \right) \Rightarrow \mathbbm{1} \otimes Z, 
\label{eq:se2}
    \\
    \left( F_{[12]}^{1,1} - F_{[12]}^{1,r^{2}} - F_{[12]}^{1,rs} + F_{[12]}^{1,r^{3}s}  \right) \Rightarrow Z \otimes Z,
\label{eq:se1e2}
\end{align} 
which take the same form as the short string operators of $e_{2}$ and $e_{1}e_{2}$ in the $\z_{2}^{2}$ surface code, respectively. Therefore, we find that the short ribbon operator $F_{\tau}^{m_{B}}$ becomes a short string operator creating a pair of composite anyons, which are equal weighted superpositions of the $e_{2}$ and $e_{1}e_{2}$ anyons. We can then use the gluing formula Eq.~\eqref{eq:gluingribbon} to build the corresponding logical operators. From the mapping of the logical operators, we conclude that the logical state after the anyon condensation becomes
\begin{equation}
    \ket{\boldsymbol{1}} + e^{i \pi/4} \ket{e_{2}} + e^{i \pi/4} \ket{e_{1}e_{2}} - \ket{e_{1}}.
    \label{eq:z22_state}
\end{equation}

\subsection{Extract a magic state in $\z_{2}$ surface code}

\subsubsection{Option (1): disentangle the logical state}
\label{sec:disentangle_tc}

Treating the code subspace of the $\z_{2}^{2}$ surface code as two logical qubits, the state in Eq.~(\ref{eq:z22_state}) can be seen as obtained from a logical entangling gate through 
\be
    (\bar{H} \otimes \bar{H}) \overline{CZ} \ket{-} \otimes \ket{T},
\ee
where $\ket{-}=\ket{e_1}$ and $\ket{T}=T\ket{+}=T\ket{\boldsymbol{1}}$. 
We can then apply two logical $\bar{H}$ gates and a logical $\overline{CZ}$ gate , which are fold-transversal or transversal, to extract the magic state $\ket{T}$ encoded by one $\z_{2}$ surface code. To see how these logical gates can be implemented, we first turn the $\z_{2}^{2}$ surface code into two decoupled $\z_{2}$ surface codes. The anyon condensation described by Eq.~(\ref{eq:P_eG}) can be realized by first applying a controlled-$\tilde{X}$ gate on each edge $l$
\be
    U_{C\tilde{X}} = \prod_{l} C\tilde{X}_{l},
    \label{eq:edge_cx}
\ee
where the control and target are the qubit and qudit, respectively, followed by measurements and post-processing that fulfill the projection
\be
    P_{l}^{\tilde{Z}^2} = \frac{1+\tilde{Z}_{l}^{2}}{2}
    \label{eq:P_z2}
\ee
for all edges. This procedure transforms the original basis states $\{\ket{0,0}, \ket{0,2}, \ket{1,3}, \ket{1,1}\}$ of the local Hilbert space at each edge, labeled by the qubit and qudit computational bases, into the basis states $\{\ket{0,0}, \ket{0,2}, \ket{1,0}, \ket{1,2}\}$, which are identified as the basis in Eq.~(\ref{eq:z22subspace}). In the new basis, the $\z_{2}^{2}$ surface code can be seen as two decoupled $\z_{2}$ surface codes supported on the edge qubits and qudits, respectively. In particular, the states $\ket{0}, \ket{2}$ of each qudit define an effective qubit, on which the qudit operators $\tilde{X}^2, \tilde{Z}$ act like qubit Pauli $X, Z$. Recall that the boundary conditions of the original qudit lattice and the added qubit lattice are opposite of each other and consequently the logical $\bar{X}$ ($\bar{Z}$) of the first copy of $\z_{2}$ surface code and the logical $\bar{Z}$ ($\bar{X}$) of the second copy have the same orientation. To apply a logical $\overline{CZ}$ gate that disentangles the two $\z_{2}$ surface codes, we can apply the following transversal operation
\be
    U_{CZ} = \prod_{l} C\tilde{Z}_{l},
\ee
where the control and target are the qubit and qudit, respectively.

\subsubsection{Option (2): condense $\mathcal{A}_{2}'=1\oplus m_{1}e_{2}$}
\label{sec:condensed_z2_surface_code}

To go from the $\z_{2}^2$ surface code to the $\z_{2}$ surface code through anyon condensation, we need to condense $m_{1}e_{2}$ particles by measuring the corresponding short ribbon operators and post-processing. There are different orientations for the short ribbon operators $F_{\tau}^{m_{R};(1,1),(1,1)}$ of the $m_{R}$ particle in the $D_{4}$ quantum double model on an edge $l$~\cite{Ellison2022}. They correspond to different ways of binding a pair of $m_1$ and $e_2$ particles together in the intermediate $\z_{2}^{2}$ toric code. We choose a set of orientations in which the short ribbon operators commute with each other. They are
\begin{align}
    C_{l}^{(1)} &= \includegraphics[width=.23\linewidth,valign=c]{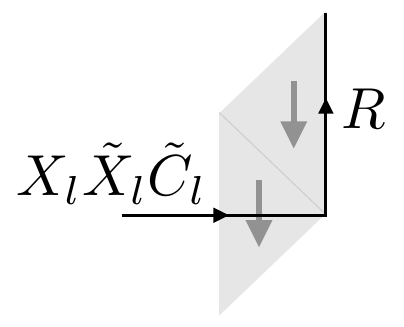}, \quad \includegraphics[width=.25\linewidth,valign=c]{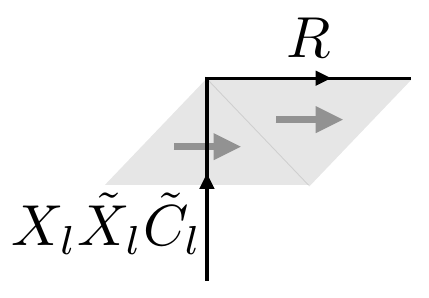}, \nonumber
    \\
    C_{l}^{(2)} &= \includegraphics[width=.2\linewidth,valign=c]{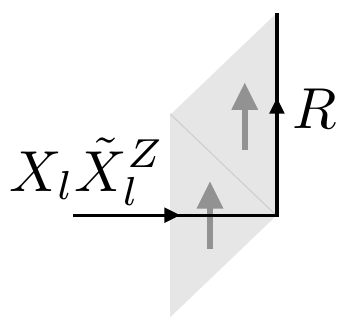}, \quad \includegraphics[width=.24\linewidth,valign=c]{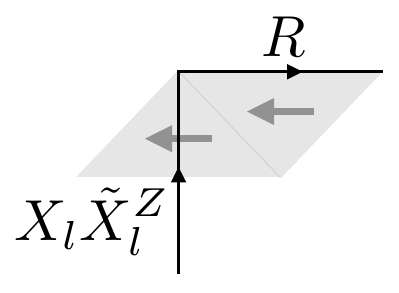},
\label{eq:sm1e2}
\end{align}
where 
\begin{align}
    R &\equiv T^{1}_+ - T^{r^{2}}_+ + T^{rs}_+ - T^{r^{3}s}_+ \nonumber
    \\
    &= \tilde{Z} \left( \frac{1+\tilde{Z}^{2}}{2} \right) \left( \frac{1+Z}{2} \right) \nonumber\\
    &+ i \tilde{Z}^{3} \left( \frac{1-\tilde{Z}^{2}}{2} \right) \left( \frac{1-Z}{2} \right).
\label{eq:R}
\end{align}
After the gates and measurements described by Eqs.~(\ref{eq:edge_cx}) and (\ref{eq:P_z2}), $R$ acts as $\mathbbm{1}\otimes Z$.

Note that the expressions of $C_{l}^{(1)}$ and $C_{l}^{(2)}$ depend on whether the edge $l$ is vertical or horizontal. These terms mutually commute. In the subspace of Eq.~\eqref{eq:z22subspace}, both operators become the same
\begin{align}
    C_{l}^{(1)} &\Rightarrow \includegraphics[width=.22\linewidth,valign=c]{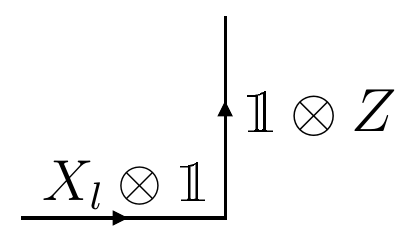}, \quad \includegraphics[width=.22\linewidth,valign=c]{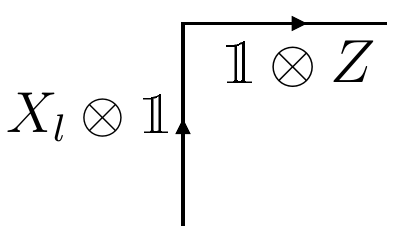}, \nonumber
    \\
    C_{l}^{(2)} &\Rightarrow \includegraphics[width=.22\linewidth,valign=c]{sm1e2_p1.pdf}, \quad \includegraphics[width=.22\linewidth,valign=c]{sm1e2_p2.pdf},
\label{eq:sm1e2_projected}
\end{align}
which is the standard form of the short $m_{1}e_{2}$ string operator in the $\z_{2}^{2}$ surface code.  

The Hamiltonian after condensation is given by
\begin{equation}
    A'_{v} =  \includegraphics[width=.24\linewidth,valign=c]{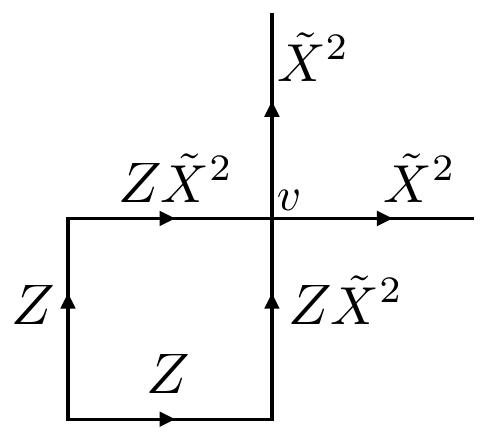}, \quad
    B'_{p} =  \includegraphics[width=.24\linewidth,valign=c]{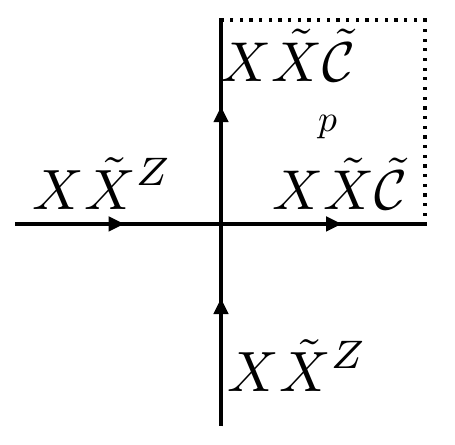}.
\end{equation}
The boundary Hamiltonian terms are presented in Appendix~\ref{app:bdy_cz2}. In the subspace Eq.~\eqref{eq:z22subspace}, the Hamiltonian terms take the form
\begin{equation}
    A'_{v} \Rightarrow \includegraphics[width=.26\linewidth,valign=c]{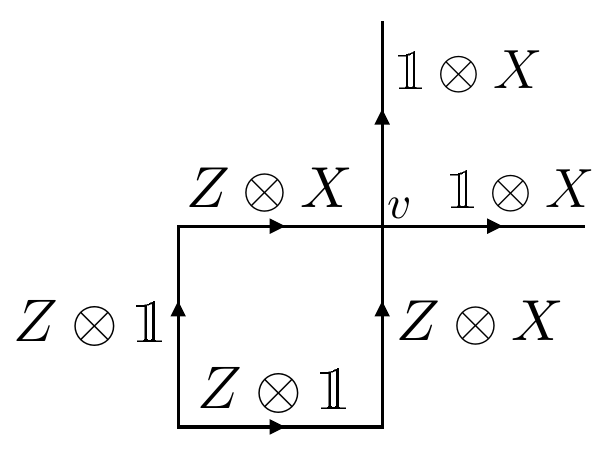}, \quad B'_{p} \Rightarrow \includegraphics[width=.26\linewidth,valign=c]{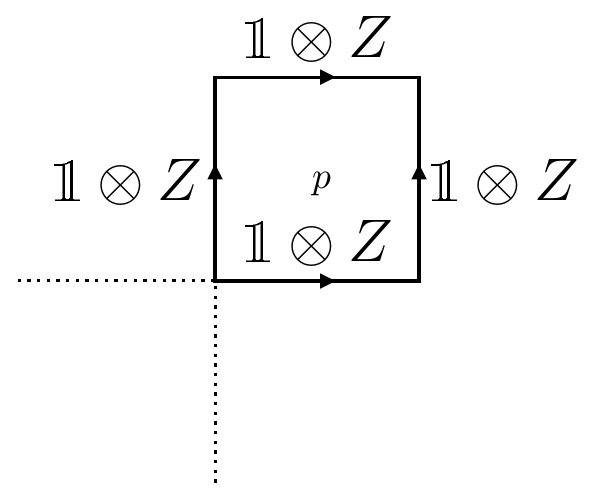}.
\end{equation}
In the subspace after condensation, this model realizes the same $\mathcal{Z}(\z_{2})$ topological order as the standard $\z_{2}$ toric code, which can be seen from the analysis of string operators below. We call this model the condensed $\z_{2}$ surface code. We can transform the condensed $\z_{2}$ surface code to the standard form by the following unitary in this subspace acting on each edge:
\be
    \includegraphics[width=.26\linewidth,valign=c]{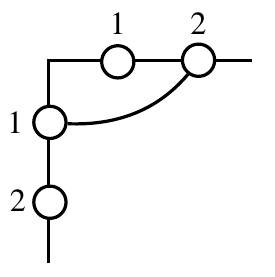}, \quad \includegraphics[width=.26\linewidth,valign=c]{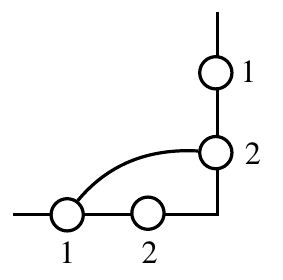}
    \label{eq:cond_to_stand}
\ee
where the curved solid line represents a $CZ$ gate. The two qubits labeled $1$ and $2$ represent the qubit and the effective qubit formed by the $\{\ket{0}, \ket{2}\}$ states of the qudit, respectively. Afterwards, the vertex and plaquette terms take the standard form supported by the second (effective) qubits on the edges. For the purpose of gate teleportation, one can either stay with the condensed $\z_{2}$ surface code or transform into the standard form first.

To see how the logical information transforms, we examine the particle string operators in the intermediate $\z_2^2$ surface code. 
It is easy to see their commutation relations in the transformed basis defined by Eq.~(\ref{eq:z22subspace}). 
In this basis, the string operators $Z\otimes\mathbbm{1}$ and $Z\otimes Z$ that create $e_{1}$ and $e_{1}e_{2}$ particles, respectively, do not commute with the ribbon operators in Eq.~\eqref{eq:sm1e2_projected}. The corresponding logical operators $L_{e_{1}}$ and $L_{e_{1}e_{2}}$ do not survive anyon condensation. 
On the other hand, the $e_{2}$ particle string $\mathbbm{1}\otimes Z$ commutes with the ribbon operators in Eq.~\eqref{eq:sm1e2_projected}, 
and becomes a string operator for the $e$ particle in the condensed $\z_{2}$ surface code. It takes the form
\begin{equation}
    S^{(e)}_{\gamma} = \prod_{l \in \gamma} R_{l} \Rightarrow \prod_{l \in \gamma} \mathbbm{1}\otimes Z_{l},
\end{equation}
where $\gamma$ is a path and the $R_{l}$ operator on a edge $l$ is defined in Eq.~\eqref{eq:R} and becomes $\mathbbm{1}\otimes Z_l$ in the transformed basis. 
Similarly, an $m$ string operator takes the form
\begin{equation}
    S^{(m)}_{\gamma} = \prod_{l \in \gamma} W^{(m)}_{l},
\label{eq:sm}
\end{equation}
where a short string operator $W^{(m)}_{l}$ is represented pictorially as
\begin{align}
    W^{(m)}_{l} &= \includegraphics[width=.22\linewidth,valign=c]{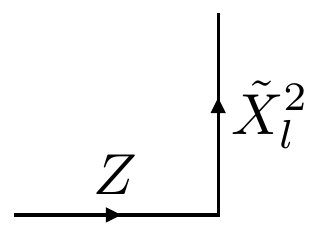}, \quad \includegraphics[width=.22\linewidth,valign=c]{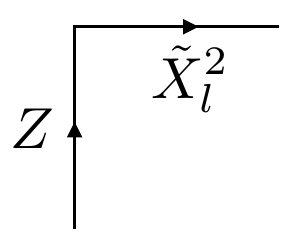} \nonumber\\
    &\Rightarrow
    \includegraphics[width=.3\linewidth,valign=c]{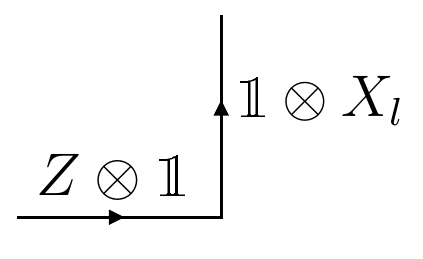}, \quad \includegraphics[width=.3\linewidth,valign=c]{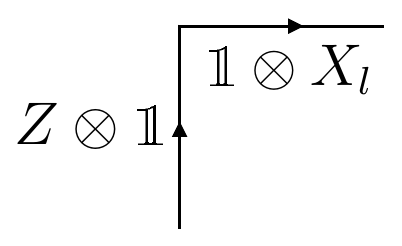}.
\end{align} 
The local gates described by Eq.~(\ref{eq:cond_to_stand}) takes the condensed $\z_{2}$ surface code to the standard form, after which $R_l\Rightarrow(\mathbbm{1}\otimes Z)_l$ stays unchanged while $W^{(m)}_{l}$ becomes $(\mathbbm{1} \otimes X)_l$. These are the short string operators for the $e$ and $m$ particles, respectively, in the standard $\z_{2}$ surface code. When extended to a pair of opposite boundaries, the string operators $S^{(e)}_{\gamma}$ and $S^{(m)}_{\gamma}$ form the logical operators $L_{e}$ and $L_{m}$, respectively. Since $\{ R,\tilde{X}^{2} \} = 0$, $L_{e}$ and $L_{m}$ logical operators anti-commute and there is a 2 dimensional logical space. $L_{e}$ can be regarded as the logical $Z$ operator and $L_{m}$ as the logical $X$ operator. Denoting the eigenstates of $L_{m}$ with the $+1$ and $-1$ eigenvalues as $\ket{\boldsymbol{1}}$ and $\ket{e}$ respectively, the logical state after anyon condensation becomes
\begin{equation}
    \ket{T_{X}} = \ket{\boldsymbol{1}} + e^{i \pi/4} \ket{e}.
\end{equation}
This logical state takes the form of the magic state, and we can use it to perform the $T$ gate via gate teleportation.

\section{$\mathrm{T}$ gate through gate teleportation}
\label{sec:tgate}

\begin{figure}
	\centering
	\includegraphics[width=0.45\textwidth]{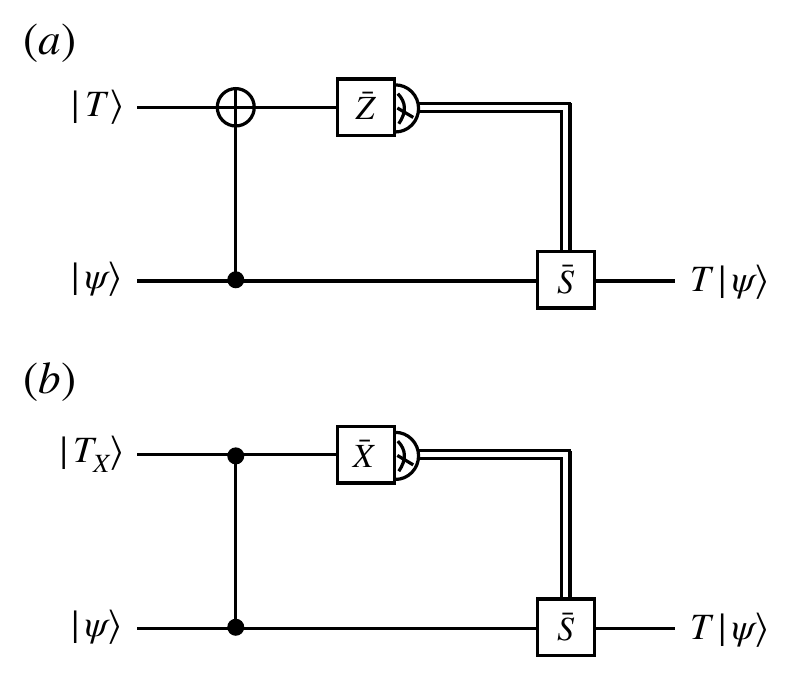}
	\caption{Gate-teleportation circuits to perform a $T$ gate by consuming (a) $\ket{T}=T\ket{\boldsymbol{1}}$ or (b) $\ket{T_{X}} = \ket{\boldsymbol{1}} + e^{i \pi/4} \ket{e}$.}
	\label{fig:t-gate_teleportation}
\end{figure}

\begin{figure}
	\centering
	\includegraphics[width=0.2\textwidth]{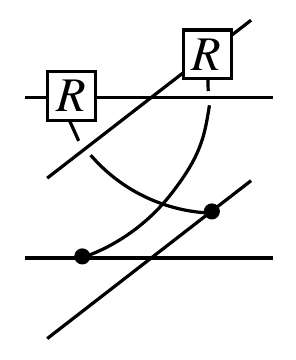}
	\caption{The logical $\overline{CZ}$ gate is performed by applying a product of physical $CR$ gates. The top layer is the condensed $\z_{2}$ surface code block and the bottom layer is the standard $\z_{2}$ surface code block.}
	\label{fig:logicalCZ}
\end{figure}

Depending on the logical $\ket{T}$ or $\ket{T_{X}}$ state, the logical teleportation circuit used for gate teleportation is shown in Fig.~\ref{fig:t-gate_teleportation}, where $\ket{\psi}$ is a logical state in a standard $\z_{2}$ surface code. If $\ket{T}$ or $\ket{T_{X}}$ is already encoded in the standard $\z_{2}$ surface code, the logical gates in the circuit take the standard transversal form. If the magic state is $\ket{T_{X}}$ in the condensed $\z_{2}$ surface code, the first step is to perform the logical $\overline{CZ}$ gate between the standard $\z_{2}$ surface code block and the condensed $\z_{2}$ surface code block. We define a physical $CR$ gate in analogy to the physical $CZ$ gate:
\begin{equation}
   CR_{j,k} = \ket{0}\bra{0}_{j} \otimes \mathbbm{1}_{k} + \ket{1}\bra{1}_{j} \otimes R_{k},
\end{equation}
where the control qubit $j$ is in the standard $\z_{2}$ surface code block and the target qubit $k$ is in the condensed $\z_{2}$ surface code block. Since the logical $L_e$ operator is a string of physical $R$ operators, the logical $\overline{CZ}$ on the 2-qubit system can be achieved by transversally applying physical $CR$ gates in a pattern shown in Fig.~\ref{fig:logicalCZ}. 
Next, the $\bar{X}$ logical operator in the condensed $\z_{2}$ surface code is measured. Depending on the measurement outcome, a classically controlled $\bar{S}$ gate is applied on the $\z_{2}$ surface code block to complete the gate teleportation. The $\bar{S}$ gate can be performed by the fold-transversal method~\cite{Moussa2016Fold} or the method in Ref.~\cite{Gidney2024Ybasis}. 

\section{Discussion}
\label{sec:discussion}

In this work, we explore a unique way of generating the logical magic state with the aid of a non-Abelian topological order. To design the procedure, we employ a framework where QEC code blocks are represented by the topological orders they realize. Under this framework, all the operations are regarded and analyzed as topological manipulations, including gauging transformation and anyon condensation. This continuum discussion, without any detail about lattices, provides a clear and global view, which helps to engineer the operations that can achieve the desired transformation on the logical information. On lattice models, i.e. the specific QEC codes, these topological manipulations can be translated back into physical operations, including adding auxiliary degrees of freedom, applying physical quantum gates, measuring a subsystem and post-processing after measurement. 

In our method, the starting point is a logical state $\ket{S_{X}}$ in the $\z_4$ surface code obtained by conjugating the logical phase gate $\bar{S}$ with the logical discrete Fourier transform gate $\bar{H}$. The goal is to prepare a logical magic state or $T$ gate in the $\z_2$ surface code. 
We now describe the method first using the language of topological order, and then in terms of the lattice models and physical operations. We start from the logical state $\ket{S_{X}}$ in an abstract code patch realizing the $\mathcal{Z}(\z_{4})$ topological order, and gauge the charge conjugation symmetry to obtain an intermediate $\mathcal{Z}(D_{4})$ topological order. 
Then, anyon condensation is performed to obtain either $\mathcal{Z}(\z_{2}^2)$ or $\mathcal{Z}(\z_{2})$, which can be seen as two logical qubits and one logical qubit, respectively. In the former, the original $\ket{S_{X}}$ is transformed into a logical state, from which a magic state can be extracted using transversal gates. In the latter, the final state is a magic state. 
To realize this procedure on the lattice, an adaptive finite-depth local-unitary circuit can be applied on $\ket{S_{X}}$ in the $\z_4$ surface code to gauge the charge conjugation symmetry, after which the code becomes the $D_{4}$ surface code (the $D_{4}$ quantum double model with certain boundary conditions). We formulate the $\z_4$ surface code with local physical degrees of freedom as $4$-dimensional qudits, and ancillary qubits are introduced during the gauging operation. One can equivalently replace each $4$-dimensional qudit with two qubits so that both $\z_4$ and $D_{4}$ surface codes can be written in terms of qubits~\cite{Ellison2022,Iqbal2024D4}. 
Subsequently, we apply anyon condensation via measurements and post-processing to obtain a commuting projector model. We provide two options of extracting a magic state. The first is applying transversal gates to disentangle the two logical qubits encoded by the $\z_2^2$ surface code, obtaining a magic state in one logical qubit. The other is further condensing another anyon to reach the $\z_{2}$ surface code, either in a rotated basis or the standard form, which encodes a magic state. With this magic state, one can implement a $T$ gate in a standard $\mathbb{Z}_2$ surface code block through gate teleportation.

Our method provides a new candidate protocol for realizing the logical magic state and $T$ gate, with the potential to be fault-tolerant. Combining magic state preparation and code transformation, the procedure avoids the resource-intensive magic state distillation and all the logical gates involved are Clifford, which are desirable features in quantum computation. In addition, our construction demonstrates the utility of designing magic state generation under the framework of topological order and topological manipulations. The same framework can be applied to analyzing transformations between topological QEC codes.  

Our method also raises several interesting questions regarding both practical applications and theoretical aspects. 
The most pressing question is whether our method can be made fault tolerant. While we are optimistic about the answer, given that quantum double models are QEC codes~\cite{Wootton2014,Cui2020kitaevsquantum}, our work has not established the fault tolerance of this method. In particular, a practical implementation of anyon condensation, syndrome measurements and correction operations throughout the procedure, and a resource analysis are needed. We leave these for future work. 
Furthermore, we can take the same topological manipulations in our method and interpret them as different lattice models and operations. One may need to select the most experimentally accessible models. For example, the standard $D_4$ quantum double model is chosen as the intermediate stage with the non-Abelian topological order. A qubit non-stabilizer model that realizes the same $\mathcal{Z}(D_{4})$ topological order~\cite{Tantivasadakarn2023nonabelian} has been prepared on a trapped-ion processor~\cite{Iqbal2024D4}. Engineering the operations corresponding to the topological manipulations on such a model would open up possibilities of demonstrating our method in experiments. 

Theoretically, the choice of intermediate topological orders (i.e. the intermediate QEC codes) in this work is determined through inspection and specifically tailored to the initial state $\ket{S_{X}}$ and the final state $\ket{T_{X}}$. A key question is whether general conditions can be established for selecting intermediate topological orders that helps transform a logical stabilizer state into a magic state. 
Additionally, it would be useful to investigate what logical operations can be achieved by the `sandwich' method described in Sec.~\ref{sec:magic_cat}. This can provide insights into the broader potential of topological manipulations in quantum computation.


\begin{acknowledgements}
The work of Sheng-Jie Huang is supported by the UKRI Frontier Research Grant, underwriting the ERC Advanced Grant ``Generalized Symmetries in Quantum Field Theory and Quantum Gravity". We thank Sakura Schafer-Nameki and Linnea Grans-Samuelsson for inspiring discussions. 

During the completion of this work, we became aware of
a related work on implementing non-Clifford gates by switching to a non-Abelian topological code~\cite{Margarita2025}.

\end{acknowledgements}


\appendix

\section{Examples of ribbon operators}
\label{app:D4_model_ribbon}
In this section, we give some examples of ribbon operators in the $D_{4}$ surface code by calculated from Eq.~\eqref{eq:ribbon_general}. These expressions are used in analyzing how the logical information transforms on the lattice models.

The ribbon operator for $e_{R}$ particle is given by
\begin{align}
    F_{\rho}^{e_{R}} &= \frac{1}{8}(F_{\rho}^{1,1}-F_{\rho}^{1,r}+F_{\rho}^{1,r^{2}}-F_{\rho}^{1,r^{3}} \nonumber\\
    &+F_{\rho}^{1,s}+F_{\rho}^{1,r^{2}s}-F_{\rho}^{1,rs}-F_{\rho}^{1,r^{3}s}),
\end{align}
where the $(\boldsymbol{u},\boldsymbol{v})$ superscript in Eq.~\eqref{eq:ribbon_general} is suppressed because the internal space of the anyon is 1-dimensional. We define $E_{\rho}$ to be the set of long edges of the direct triangles in the ribbon $\rho$. Using Eq.~\eqref{fig:ribbon_action}, we can write the ribbon operator as 
\begin{equation}
    F_{\rho}^{e_{R}} = \prod_{l \in E_{\rho}} \tilde{Z}^{2}_{l}.
\end{equation}

The ribbon operator for $e_{G}$ particle is given by
\begin{align}
    F_{\rho}^{e_{G}} &= \frac{1}{8}(F_{\rho}^{1,1}-F_{\rho}^{1,r}+F_{\rho}^{1,r^{2}}-F_{\rho}^{1,r^{3}} \nonumber\\
    &-F_{\rho}^{1,s}+F_{\rho}^{1,r^{2}s}+F_{\rho}^{1,rs}+F_{\rho}^{1,r^{3}s}).
\end{align}
Using Eq.~\eqref{fig:ribbon_action}, the ribbon operator becomes 
\begin{equation}
    F_{\rho}^{e_{G}} = \prod_{l \in E_{\rho}} \tilde{Z}^{2}_{l}Z_{l}.
\end{equation}

The ribbon operators of $m_{B}$ particle are
\begin{align}
    F_{\rho}^{m_{B};((1,1),(1,1))} &= \frac{1}{4}\left( F_{\rho}^{1,1} - i F_{\rho}^{1,r} - F_{\rho}^{1,r^{2}} + i F_{\rho}^{1,r^{3}} \right), \nonumber
    \\
    F_{\rho}^{m_{B};((1,2),(1,2))} &= \frac{1}{4}\left( F_{\rho}^{1,1} + i F_{\rho}^{1,r} - F_{\rho}^{1,r^{2}} - i F_{\rho}^{1,r^{3}} \right), \nonumber
    \\
    F_{\rho}^{m_{B};((1,1),(1,2))} &= \frac{1}{4}\left( F_{\rho}^{1,s} + i F_{\rho}^{1,rs} - F_{\rho}^{1,r^{2}s} - i F_{\rho}^{1,r^{3}s} \right), \nonumber
    \\
    F_{\rho}^{m_{B};((1,2),(1,1))} &= \frac{1}{4}\left( F_{\rho}^{1,s} - i F_{\rho}^{1,rs} - F_{\rho}^{1,r^{2}s} + i F_{\rho}^{1,r^{3}s} \right).
\label{eq:mb_ribbons}
\end{align}

Since the ribbon in Eq.~\eqref{eq:mb_ribbons} only involves $F_{\rho}^{1,g}$, which acts as an identity operator on all the dual triangles, the ribbon operator can written as a sum of projectors. The expression can be simplified for a specific ribbon. Let's denote the group element of $D_4$ as $g(p,q) = r^{p}s^{q}$. Consider a ribbon whose orientation coincides with the orientations of the edges of the lattice, the ribbon operators can be written as
\begin{align}
    F_{\rho}^{m_{B};((1,1),(1,1))} &= \sum_{p=0}^{3}\left( 
    i^{-p} P_{\rho}^{g(p,0)},
    \right), \nonumber
    \\
    F_{\rho}^{m_{B};((1,2),(1,2))} &= \sum_{p=0}^{3}\left( 
    i^{p} P_{\rho}^{g(p,0)},
    \right), \nonumber
    \\
    F_{\rho}^{m_{B};((1,1),(1,2))}  &= \sum_{p=0}^{3}\left( 
    i^{p} P_{\rho}^{g(p,1)}
    \right), \nonumber
    \\
    F_{\rho}^{m_{B};((1,2),(1,1))} &= \sum_{p=0}^{3}\left( 
    i^{-p} P_{\rho}^{g(p,1)}
    \right),
\label{eq:mb_ribbons_proj}
\end{align}
where $P_{\rho}^{g(p,q)} = \delta_{\prod_{l \in E_{\rho}}g_{l},g}$. 
Any basis transformation within the internal space $(\boldsymbol{u},\boldsymbol{v})$ of the anyon does not change the type of the logical operator. In particular, the combination $\text{Tr}[F_{\rho}^{m_{B};(\boldsymbol{u},\boldsymbol{v})}] + \text{ATr}[F_{\rho}^{m_{B};(\boldsymbol{u},\boldsymbol{v})}]$ in Eq.~\eqref{eq:mb_tratr} is a logical operator.

\section{Anyon condensation and Lagrangian algebra}
\label{app:anyon_condensation}
In this appendix, we give a brief review of the theory of anyon condensation and the Lagrangian algebra in a MTC $\mathcal{C}$~\cite{Kong2014,Lou2021}. An algebra in a MTC $\mathcal{C}$ is a direct sum of simple objects: $\mathcal{A} = \oplus_{\alpha} w_{\alpha} \alpha$, $\alpha \in \mathcal{C}$. More formally, we define a Lagrangian algebra $\mathcal{A}$ in a MTC $\mathcal{C}$ to be an object $\mathcal{A} \in \mathcal{C}$ along with a multiplication morphism $\mu : \mathcal{A} \otimes \mathcal{A} \rightarrow \mathcal{A}$ and unit morphism $\iota: 1 \rightarrow \mathcal{A}$ such that the following conditions hold:
\begin{itemize}
    \item Commutativity: $\mu \circ P_{\cal{A},\cal{A}} = \mu$, where $P_{\cal{A},\cal{A}}$ is the braiding in $\mathcal{C}$. It can be expressed diagrammatically:
    \begin{equation}
    \includegraphics[width=.35\linewidth]{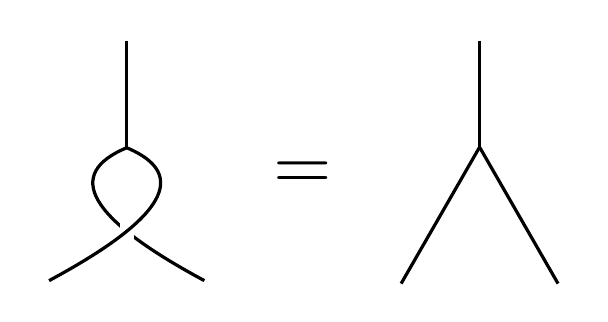},
    \end{equation}
    where solid line represents $\mathcal{A}$ and the junction where three $\mathcal{A}$ lines meet is the morphism $\mu$.
    
    \item Associativity: $\mu \circ (\mu \otimes id) = \mu \circ (id \otimes \mu)$,
    \begin{equation}
    \includegraphics[width=.4\linewidth]{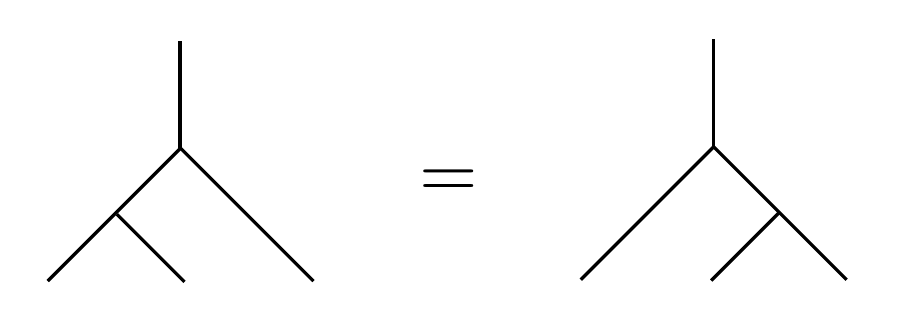}.
    \end{equation}

    \item Unit: $\mu \circ (\iota \otimes id) = id$,
    \begin{equation}
    \includegraphics[width=.2\linewidth]{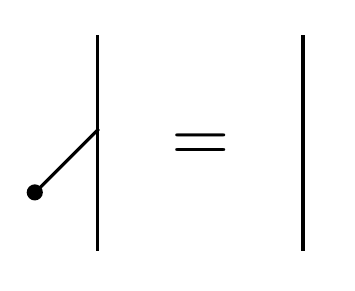}.
    \end{equation}
    $\mathcal{A}$ is called connected if $\text{Hom}(1,\mathcal{A})$ is 1-dimensional

    \item Separability: There exists a splitting morphism $\sigma : \mathcal{A} \rightarrow \mathcal{A}  \otimes \mathcal{A}$ such that $\mu \circ \sigma = id$, and satisfies
    \begin{equation}
    \includegraphics[width=.55\linewidth]{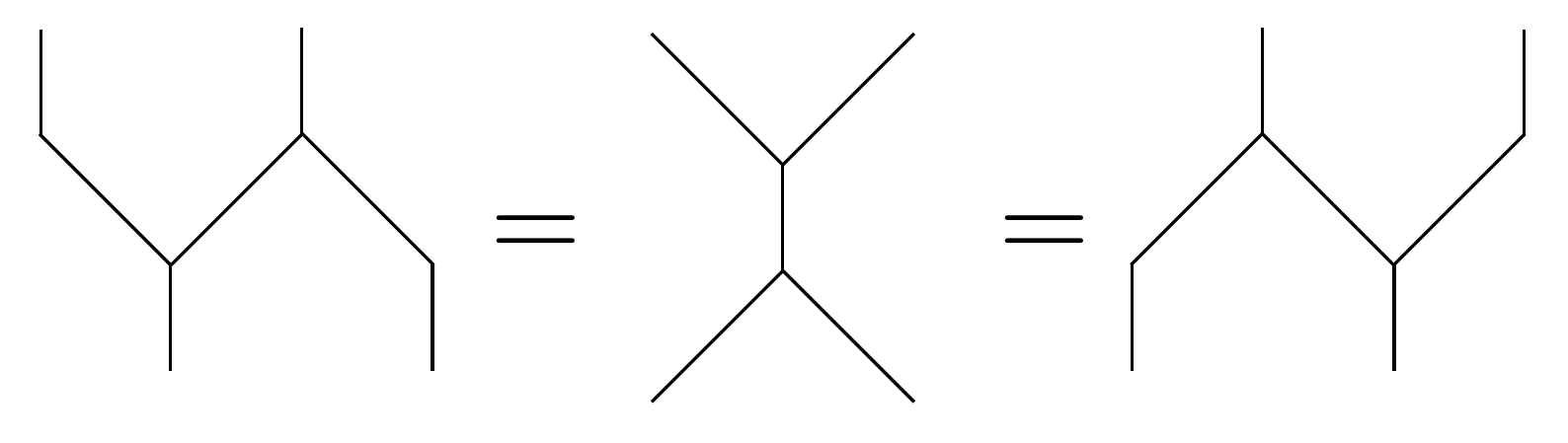}.
    \end{equation}

    \item Lagrangian:
    $D_{\mathcal{A}} = D_{\mathcal{C}}$,
    where $D_{\mathcal{A}} = \sum_{\alpha \in \mathcal{C}} w_{\alpha} d_{\alpha}$ is the dimension of the algebra $\mathcal{A} = \oplus_{\alpha} w_{\alpha} \alpha$, and $D_{\mathcal{C}} = \sqrt{\sum_{\alpha \in \mathcal{C}} d_{\alpha}^{2}}$ is the total quantum dimension of $\mathcal{C}$.
\end{itemize}

It can be shown that $\mathcal{A}$ is a commutative associative algebra in a MTC $\mathcal{C}$ if and only if $\mathcal{A}$ decomposes into simple objects as $\mathcal{A} = \oplus_{\alpha} n_{\alpha} \alpha$, with $\theta_{\alpha} = 1$ for all $\alpha$ such that $n_{\alpha} \neq 0$~\cite{Frohlich2006}. Physically, it means that we enter a trivially gapped phase after the anyon condensation given by $\mathcal{A}$, or equivalently $\mathcal{A}$ describes the set of anyons that are condensed. 

When the Lagrangian condition is not satisfied, the algebra $\mathcal{A}$ is called a \emph{condensable algebra}. The resulting phase after the anyon condensation is a reduced topological order
\begin{equation}
    \mathcal{Z'} = \mathcal{Z}/\mathcal{A}.
\end{equation}
It is possible to write the necessary conditions for a condensable algebra $\mathcal{A}$ in terms of $n_{\alpha}$. The sufficient condition however requires knowledge of the reduced topological order. We are usually interested in the mapping between the anyons in the two theories. This is described in a restriction map
\begin{equation}
    a \rightarrow \bigoplus_{a' \in \mathcal{Z'}} n_{a,a'} a',
\label{eq:restriction}
\end{equation}
which expresses an anyon $a \in \mathcal{Z}$ as a formal sum of
anyons $a' \in \mathcal{Z'}$, and a lift 
\begin{equation}
    a' \rightarrow \bigoplus_{a \in \mathcal{Z}} n_{a,a'} a.
\label{eq:lift}
\end{equation}
For example, the vacuum $1$ of the new theory should lift to the condensable algebra $\mathcal{A}$. See Ref.~\cite{Chatterjee2023Holo,Bhardwaj2023Club} for more discussion. 

Here we list the lifts that are used in this work. The lift $\mathcal{Z}(\z_{4}) \rightarrow \mathcal{Z}(D_{4})$ for the condensable algebra $\mathcal{A} = 1 \oplus e_{RG}$ is\footnote{The algebra is labeled as $\mathcal{A}_{1}$ in Ref.~\cite{Bhardwaj2024Hasse}. We use the lift provided in Ref.~\cite{Bhardwaj2024Hasse} with $e$ and $m$ exchanged.}
\begin{gather}
    1 \rightarrow  1 \oplus e_{RG}, \quad e \rightarrow m_{B}, \quad e^{2} \rightarrow e_{R} \oplus e_{G}, \nonumber\\
    e^{3} \rightarrow m_{B}, \quad m \rightarrow m_{RG}, \quad m^{2} \rightarrow e_{RGB} \oplus e_{B}, \nonumber\\
    m^{3} \rightarrow m_{RG}, \quad e^{2}m^{2} \rightarrow e_{GB} \oplus e_{RB}, \quad e^{2}m \rightarrow f_{B}, \nonumber\\
    e^{2}m^{3} \rightarrow f_{B}, \quad em^{2} \rightarrow f_{RG}, \quad e^{3}m^{2} \rightarrow f_{RG}, \nonumber\\
    em  \rightarrow s_{RGB}, \quad e^{3}m^{3}  \rightarrow s_{RGB}, \quad e^{3}m  \rightarrow \bar{s}_{RGB}, \nonumber\\
    em^{3}  \rightarrow \bar{s}_{RGB}.
\label{eq:z4liftd4}
\end{gather}

The lift $\mathcal{Z}(\z_{2}) \rightarrow \mathcal{Z}(D_{4})$ for the algebra $\mathcal{A}' = 1 \oplus e_{G} \oplus m_{R}$ is given by
\begin{align}
    &1 \rightarrow 1 \oplus e_{G} \oplus m_{R}, \nonumber\\ 
    &e \rightarrow m_{B} \oplus m_{RB}, \nonumber\\ 
    &m \rightarrow e_{B} \oplus e_{GB} \oplus m_{R}, \nonumber\\
    &f \rightarrow f_{B} \oplus f_{RB}.
\end{align}

The lift $\mathcal{Z}(\z_{2}^{2}) \rightarrow \mathcal{Z}(D_{4})$ for the algebra $\mathcal{A}_{1}' = 1 \oplus e_{G}$ is given by\footnote{The algebra is labeled as $\mathcal{A}_{5}$ in Ref.~\cite{Bhardwaj2024Hasse}. We have used the automorphism $\rho$ to permute the anyons in the lift provided in Ref.~\cite{Bhardwaj2024Hasse} so that the string operators for the $\mathcal{Z}(\mathbb{Z}_{2}^{2})$ anyons take a convention form on the lattice. The automorphism $\rho$ we used is given by $\rho = \sigma_{s} \circ \sigma_{2} \circ \sigma_{s} \circ \sigma_{1}$, where $\sigma_{i}$ for $i=1,2$ denotes the automorphism that exchanges $e_{i} \leftrightarrow m_{i}$. The automorphism $\sigma_{s}$ implements the anyon permutation $m_{1} \rightarrow m_{1}e_{2}$, $m_{1} \rightarrow m_{2}e_{1}$, and leaves $e_{i}$ unchanged.}
\begin{gather}
     1 \rightarrow 1 \oplus e_{G}, \quad m_{1} \rightarrow m_{RB}, \quad m_{2} \rightarrow e_{RGB} \oplus e_{RB}, \nonumber\\
     m_{1}m_{2} \rightarrow m_{RB}, \quad e_{1} \rightarrow e_{RG} \oplus e_{R}, \quad e_{2} \rightarrow m_{B}, \nonumber\\
     e_{1}e_{2} \rightarrow m_{B}, \quad e_{1}m_{2} \rightarrow e_{B} \oplus e_{GB}, \quad m_{1}e_{2} \rightarrow m_{R}, \nonumber\\
     e_{1}m_{1}e_{2}m_{2} \rightarrow m_{R}.
\label{eq:anyonmapD4Z22}
\end{gather}

The lift $\mathcal{Z}(\z_{2}) \rightarrow \mathcal{Z}(\z_{2}^{2})$ for the algebra $1 \oplus m_{1}e_{2}$ is given by
\begin{align}
    &1 \rightarrow 1 \oplus m_{1}e_{2}, \nonumber\\
    &e \rightarrow m_{1} \oplus e_{2}, \nonumber\\
    &m \rightarrow e_{1}m_{2} \oplus f_{1}f_{2}, \nonumber\\
    &f \rightarrow e_{1}f_{2} \oplus f_{1}m_{2}. 
\end{align}

\section{Direct transformation from $\mathcal{Z}(\z_{4})$ to $\mathcal{Z}(\z_{2})$ by anyon condensation}
\label{app:z4z2}
Here we discuss transformation from $\mathcal{Z}(\z_{4})$ to $\mathcal{Z}(\z_{2})$ code by directly condensing anyons. There are two possibilities: condensing $m^2$ or $e^2$ anyons. We explain why they fail to generate the magic state $\ket{T_{X}}$. We first discuss the $m^{2}$ condensation. The resulting topological order is $\mathcal{Z}(\z_{2})$. The lift of the anyons is given by
\begin{align}
    1 \rightarrow 1 \oplus m^{2}, \quad e' \rightarrow e^{2} \oplus e^{2}m^{2}, \quad m' \rightarrow m \oplus m^{3}.
\end{align}
We thus have the mapping of the logical operators
\begin{equation}
    L_{m} \rightarrow L_{m'}, \quad L_{m^{3}} \rightarrow L_{m'}, \quad L_{e^{2}} \rightarrow L_{e'}.
\end{equation}
The logical state $\ket{S_{X}} = \ket{\boldsymbol{1}} + e^{i \pi/4} \ket{e} - \ket{e^{2}} + e^{i \pi/4} \ket{e^{3}}$ becomes
$\ket{\boldsymbol{1}} - \ket{e'}$ since $\ket{e}$ and $\ket{e^{3}}$ are projected out of the code space during the condensation.

We then discuss condensing $e^{2}$ particle in $\mathcal{Z}(\z_{4})$ code. The lift of the anyons are given by
\begin{align}
    1 \rightarrow 1 \oplus e^{2}, \quad e' \rightarrow e \oplus e^{3}, \quad m' \rightarrow m^{2} \oplus e^{2}m^{2},
\end{align}
which gives the mapping of the logical operators
\begin{equation}
    L_{e} \rightarrow L_{e'}, \quad L_{e^{3}} \rightarrow L_{e'}, \quad L_{m^{2}} \rightarrow L_{m'}.
\end{equation}
Since the logical state $\ket{e^{2}}$ is mapped into $\ket{\boldsymbol{1}}$ in the $\mathcal{Z}(\z_{2})$ code, the logical state $\ket{S_{X}} = \ket{\boldsymbol{1}} + e^{i \pi/4} \ket{e} - \ket{e^{2}} + e^{i \pi/4} \ket{e^{3}}$ actually becomes $0$ due to the minus sign in front of $\ket{e^{2}}$, which means that $\ket{S_{X}}$ is projected out of the code subspace. 

\section{Boundary terms of the condensed $\z_{2}$ surface code}
\label{app:bdy_cz2}
Here we present the boundary Hamiltonian terms of the condensed $\z_{2}$ surface code. The thick blue lines denote the boundary edges of the qudits, and the doubled green lines denote the boundary edges of the qubits. For the top boundary, we have the boundary terms:
\begin{equation}
    A_{v}^{T} =  \includegraphics[width=.22\linewidth,valign=c]{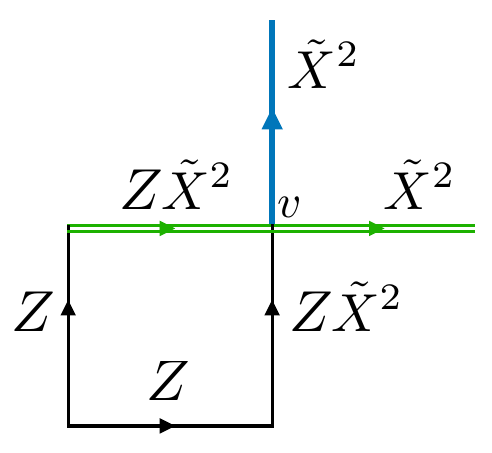}, \quad
    B_{p}^{T} =  \includegraphics[width=.22\linewidth,valign=c]{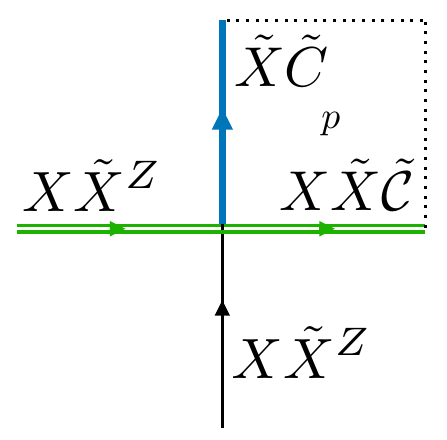}.
\end{equation}
Similarly, the boundary terms for the bottom boundary are
\begin{equation}
    A_{v}^{B} =  \includegraphics[width=.22\linewidth,valign=c]{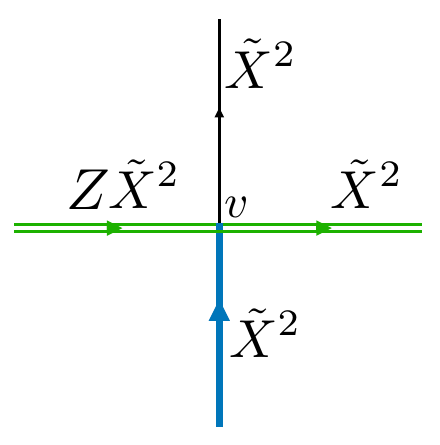}, \quad
    B_{p}^{B} =  \includegraphics[width=.22\linewidth,valign=c]{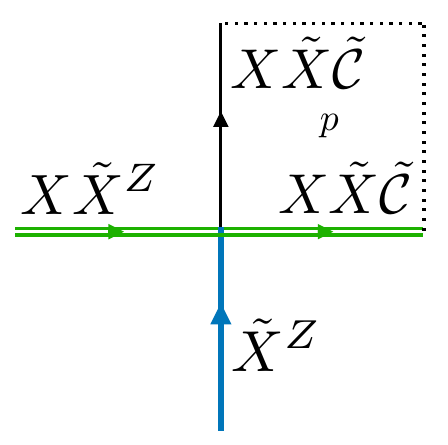}.
\end{equation}
The boundary terms for the right boundary are
\begin{equation}
    A_{v}^{R} =  \includegraphics[width=.22\linewidth,valign=c]{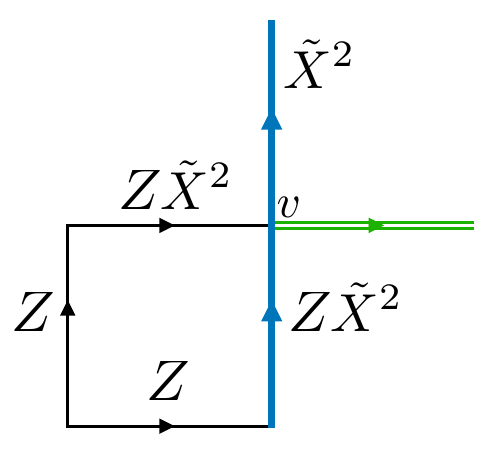}, \quad
    B_{p}^{R} =  \includegraphics[width=.22\linewidth,valign=c]{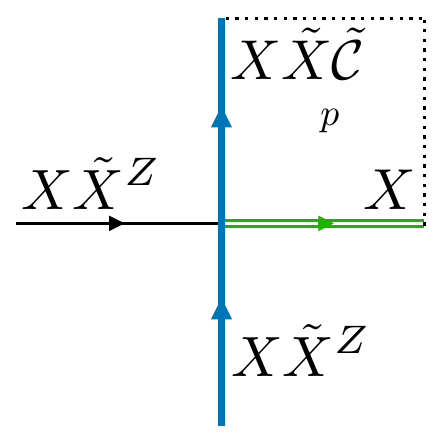},
\end{equation}
and, for the left boundary,
\begin{equation}
    A_{v}^{L} =  \includegraphics[width=.22\linewidth,valign=c]{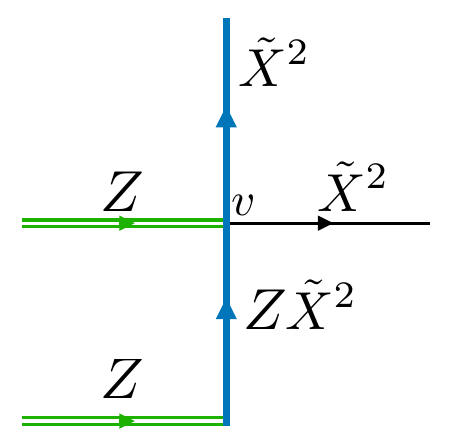}, \quad
    B_{p}^{L} =  \includegraphics[width=.22\linewidth,valign=c]{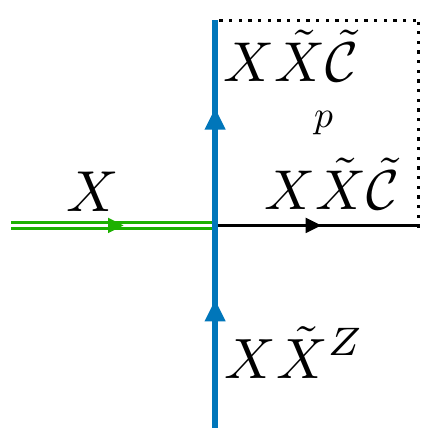}.
\end{equation}
With the basis transformation enabled by the local gates described in Sec.~\ref{sec:condensed_z2_surface_code}, these boundary terms become the rough (for the top and bottom) and smooth (for the left and right) boundaries of the standard $\z_2$ surface code.


\bibliographystyle{quantum}
\bibliography{quditsc}

\begin{thebibliography}{10}

\bibitem{Shor1995}
Peter~W. Shor.
\newblock ``Scheme for reducing decoherence in quantum computer memory''.
\newblock \href{https://dx.doi.org/10.1103/PhysRevA.52.R2493}{Phys. Rev. A {\bf 52}, R2493--R2496}~(1995).

\bibitem{Kitaev1997}
A.~Yu. Kitaev.
\newblock ``Quantum error correction with imperfect gates''.
\newblock \href{https://dx.doi.org/10.1007/978-1-4615-5923-8_19}{Pages 181--188}.
\newblock Springer US. Boston, MA~(1997).

\bibitem{Cochrane1999}
P.~T. Cochrane, G.~J. Milburn, and W.~J. Munro.
\newblock ``Macroscopically distinct quantum-superposition states as a bosonic code for amplitude damping''.
\newblock \href{https://dx.doi.org/10.1103/PhysRevA.59.2631}{Phys. Rev. A {\bf 59}, 2631--2634}~(1999).

\bibitem{Gottesman2001}
Daniel Gottesman, Alexei Kitaev, and John Preskill.
\newblock ``Encoding a qubit in an oscillator''.
\newblock \href{https://dx.doi.org/10.1103/PhysRevA.64.012310}{Phys. Rev. A {\bf 64}, 012310}~(2001).

\bibitem{Gottesman2009}
Daniel {Gottesman}.
\newblock ``{An Introduction to Quantum Error Correction and Fault-Tolerant Quantum Computation}''~(2009).
\newblock  \href{http://arxiv.org/abs/0904.2557}{arXiv:0904.2557}.

\bibitem{Dennis2002}
Eric Dennis, Alexei Kitaev, Andrew Landahl, and John Preskill.
\newblock ``Topological quantum memory''.
\newblock \href{https://dx.doi.org/10.1063/1.1499754}{J. Math. Phys. {\bf 43}, 4452}~(2002).

\bibitem{Fowler2012}
Austin~G. Fowler, Matteo Mariantoni, John~M. Martinis, and Andrew~N. Cleland.
\newblock ``Surface codes: Towards practical large-scale quantum computation''.
\newblock \href{https://dx.doi.org/10.1103/PhysRevA.86.032324}{Phys. Rev. A {\bf 86}, 032324}~(2012).

\bibitem{Horsman2012}
Dominic Horsman, Austin~G Fowler, Simon Devitt, and Rodney~Van Meter.
\newblock ``Surface code quantum computing by lattice surgery''.
\newblock \href{https://dx.doi.org/10.1088/1367-2630/14/12/123011}{New Journal of Physics {\bf 14}, 123011}~(2012).

\bibitem{Fowler2018}
Austin~G. {Fowler} and Craig {Gidney}.
\newblock ``{Low overhead quantum computation using lattice surgery}''~(2018).
\newblock  \href{http://arxiv.org/abs/1808.06709}{arXiv:1808.06709}.

\bibitem{Litinski2019gameofsurfacecodes}
Daniel Litinski.
\newblock ``A {G}ame of {S}urface {C}odes: {L}arge-{S}cale {Q}uantum {C}omputing with {L}attice {S}urgery''.
\newblock \href{https://dx.doi.org/10.22331/q-2019-03-05-128}{{Quantum} {\bf 3}, 128}~(2019).

\bibitem{Eastin2009}
Bryan Eastin and Emanuel Knill.
\newblock ``Restrictions on transversal encoded quantum gate sets''.
\newblock \href{https://dx.doi.org/10.1103/PhysRevLett.102.110502}{Phys. Rev. Lett. {\bf 102}, 110502}~(2009).

\bibitem{Brown2017}
Benjamin~J. Brown, Katharina Laubscher, Markus~S. Kesselring, and James~R. Wootton.
\newblock ``Poking holes and cutting corners to achieve clifford gates with the surface code''.
\newblock \href{https://dx.doi.org/10.1103/PhysRevX.7.021029}{Phys. Rev. X {\bf 7}, 021029}~(2017).

\bibitem{Bravyi2005magicstate}
Sergey Bravyi and Alexei Kitaev.
\newblock ``Universal quantum computation with ideal clifford gates and noisy ancillas''.
\newblock \href{https://dx.doi.org/10.1103/PhysRevA.71.022316}{Phys. Rev. A {\bf 71}, 022316}~(2005).

\bibitem{Gidney2019efficientmagicstate}
Craig Gidney and Austin~G. Fowler.
\newblock ``Efficient magic state factories with a catalyzed {$|CCZ\rangle$} to {$2|T\rangle$} transformation''.
\newblock \href{https://dx.doi.org/10.22331/q-2019-04-30-135}{{Quantum} {\bf 3}, 135}~(2019).

\bibitem{Litinski2019magicstate}
Daniel Litinski.
\newblock ``Magic {S}tate {D}istillation: {N}ot as {C}ostly as {Y}ou {T}hink''.
\newblock \href{https://dx.doi.org/10.22331/q-2019-12-02-205}{{Quantum} {\bf 3}, 205}~(2019).

\bibitem{Anderson2014}
Jonas~T. Anderson, Guillaume Duclos-Cianci, and David Poulin.
\newblock ``Fault-tolerant conversion between the steane and reed-muller quantum codes''.
\newblock \href{https://dx.doi.org/10.1103/PhysRevLett.113.080501}{Phys. Rev. Lett. {\bf 113}, 080501}~(2014).

\bibitem{Kubica2015}
Aleksander Kubica and Michael~E. Beverland.
\newblock ``Universal transversal gates with color codes: A simplified approach''.
\newblock \href{https://dx.doi.org/10.1103/PhysRevA.91.032330}{Phys. Rev. A {\bf 91}, 032330}~(2015).

\bibitem{Bombin2015}
H{\'e}ctor Bomb{\'\i}n.
\newblock ``Gauge color codes: optimal transversal gates and gauge fixing in topological stabilizer codes''.
\newblock \href{https://dx.doi.org/10.1088/1367-2630/17/8/083002}{New Journal of Physics {\bf 17}, 083002}~(2015).

\bibitem{Daguerre2024}
Lucas Daguerre and Isaac~H. Kim.
\newblock ``Code switching revisited: Low-overhead magic state preparation using color codes''.
\newblock \href{https://dx.doi.org/10.1103/PhysRevResearch.7.023080}{Phys. Rev. Res. {\bf 7}, 023080}~(2025).

\bibitem{Beverland2021}
Michael~E. Beverland, Aleksander Kubica, and Krysta~M. Svore.
\newblock ``Cost of universality: A comparative study of the overhead of state distillation and code switching with color codes''.
\newblock \href{https://dx.doi.org/10.1103/PRXQuantum.2.020341}{PRX Quantum {\bf 2}, 020341}~(2021).

\bibitem{Moussa2016}
Jonathan~E. Moussa.
\newblock ``Quantum circuits for qubit fusion''.
\newblock \href{https://dx.doi.org/https://doi.org/10.26421/QIC16.13-14-3}{Quantum Info. Comput. {\bf 16}, 1113–1124}~(2016).

\bibitem{Moussa2016Fold}
Jonathan~E. Moussa.
\newblock ``Transversal clifford gates on folded surface codes''.
\newblock \href{https://dx.doi.org/10.1103/PhysRevA.94.042316}{Phys. Rev. A {\bf 94}, 042316}~(2016).

\bibitem{Kitaev2003}
A.Yu. Kitaev.
\newblock ``Fault-tolerant quantum computation by anyons''.
\newblock \href{https://dx.doi.org/https://doi.org/10.1016/S0003-4916(02)00018-0}{Annals of Physics {\bf 303}, 2--30}~(2003).

\bibitem{Nayak2008}
Chetan Nayak, Steven~H. Simon, Ady Stern, Michael Freedman, and Sankar Das~Sarma.
\newblock ``Non-abelian anyons and topological quantum computation''.
\newblock \href{https://dx.doi.org/10.1103/RevModPhys.80.1083}{Rev. Mod. Phys. {\bf 80}, 1083--1159}~(2008).

\bibitem{wang2008}
Zhenghan Wang.
\newblock ``Topological quantum computation''.
\newblock \href{https://dx.doi.org/10.1090/cbms/112}{Volume 112 of CBMS Regional Conference Series in Mathematics}.
\newblock American Mathematical Society. Providence, RI~(2008).

\bibitem{Drinfeld2010}
Vladimir Drinfeld, Shlomo Gelaki, Dmitri Nikshych, and Victor Ostrik.
\newblock ``On braided fusion categories i''.
\newblock \href{https://dx.doi.org/10.1007/s00029-010-0017-z}{Selecta Mathematica {\bf 16}, 1--119}~(2010).

\bibitem{Haegeman2015}
Jutho Haegeman, Karel Van~Acoleyen, Norbert Schuch, J.~Ignacio Cirac, and Frank Verstraete.
\newblock ``Gauging quantum states: From global to local symmetries in many-body systems''.
\newblock \href{https://dx.doi.org/10.1103/PhysRevX.5.011024}{Phys. Rev. X {\bf 5}, 011024}~(2015).

\bibitem{Kubica2018}
Aleksander {Kubica} and Beni {Yoshida}.
\newblock ``{Ungauging quantum error-correcting codes}''~(2018).
\newblock  \href{http://arxiv.org/abs/1805.01836}{arXiv:1805.01836}.

\bibitem{Tantivasadakarn2024}
Nathanan {Tantivasadakarn}, Ryan {Thorngren}, Ashvin {Vishwanath}, and Ruben {Verresen}.
\newblock ``{Long-Range Entanglement from Measuring Symmetry-Protected Topological Phases}''.
\newblock \href{https://dx.doi.org/10.1103/PhysRevX.14.021040}{Physical Review X {\bf 14}, 021040}~(2024).
\newblock  \href{http://arxiv.org/abs/2112.01519}{arXiv:2112.01519}.

\bibitem{Tantivasadakarn2023Long}
Nathanan Tantivasadakarn, Ashvin Vishwanath, and Ruben Verresen.
\newblock ``Hierarchy of topological order from finite-depth unitaries, measurement, and feedforward''.
\newblock \href{https://dx.doi.org/10.1103/PRXQuantum.4.020339}{PRX Quantum {\bf 4}, 020339}~(2023).

\bibitem{Bais2009}
F.~A. Bais and J.~K. Slingerland.
\newblock ``Condensate-induced transitions between topologically ordered phases''.
\newblock \href{https://dx.doi.org/10.1103/PhysRevB.79.045316}{Phys. Rev. B {\bf 79}, 045316}~(2009).

\bibitem{Eliens2014}
I.~S. Eli\"ens, J.~C. Romers, and F.~A. Bais.
\newblock ``Diagrammatics for bose condensation in anyon theories''.
\newblock \href{https://dx.doi.org/10.1103/PhysRevB.90.195130}{Phys. Rev. B {\bf 90}, 195130}~(2014).

\bibitem{Kong2014}
Liang Kong.
\newblock ``Anyon condensation and tensor categories''.
\newblock \href{https://dx.doi.org/https://doi.org/10.1016/j.nuclphysb.2014.07.003}{Nuclear Physics B {\bf 886}, 436--482}~(2014).

\bibitem{Cong2017PRL}
Iris Cong, Meng Cheng, and Zhenghan Wang.
\newblock ``Universal quantum computation with gapped boundaries''.
\newblock \href{https://dx.doi.org/10.1103/PhysRevLett.119.170504}{Phys. Rev. Lett. {\bf 119}, 170504}~(2017).

\bibitem{Cong2017}
Iris Cong, Meng Cheng, and Zhenghan Wang.
\newblock ``Hamiltonian and algebraic theories of gapped boundaries in topological phases of matter''.
\newblock \href{https://dx.doi.org/10.1007/s00220-017-2960-4}{Communications in Mathematical Physics {\bf 355}, 645--689}~(2017).

\bibitem{Burnell2018}
F.~J. {Burnell}.
\newblock ``{Anyon Condensation and Its Applications}''.
\newblock \href{https://dx.doi.org/10.1146/annurev-conmatphys-033117-054154}{Annual Review of Condensed Matter Physics {\bf 9}, 307--327}~(2018).
\newblock  \href{http://arxiv.org/abs/1706.04940}{arXiv:1706.04940}.

\bibitem{Lou2021}
Jiaqi Lou, Ce~Shen, Chaoyi Chen, and Ling-Yan Hung.
\newblock ``A (dummy's) guide to working with gapped boundaries via (fermion) condensation''.
\newblock \href{https://dx.doi.org/10.1007/JHEP02(2021)171}{Journal of High Energy Physics {\bf 2021}, 171}~(2021).

\bibitem{Kesselring2024}
Markus~S. Kesselring, Julio~C. Magdalena de~la Fuente, Felix Thomsen, Jens Eisert, Stephen~D. Bartlett, and Benjamin~J. Brown.
\newblock ``Anyon condensation and the color code''.
\newblock \href{https://dx.doi.org/10.1103/PRXQuantum.5.010342}{PRX Quantum {\bf 5}, 010342}~(2024).

\bibitem{Gottesman1999}
Daniel Gottesman and Isaac~L. Chuang.
\newblock ``Demonstrating the viability of universal quantum computation using teleportation and single-qubit operations''.
\newblock \href{https://dx.doi.org/10.1038/46503}{Nature {\bf 402}, 390--393}~(1999).

\bibitem{Piroli2021}
Lorenzo Piroli, Georgios Styliaris, and J.~Ignacio Cirac.
\newblock ``Quantum circuits assisted by local operations and classical communication: Transformations and phases of matter''.
\newblock \href{https://dx.doi.org/10.1103/PhysRevLett.127.220503}{Phys. Rev. Lett. {\bf 127}, 220503}~(2021).

\bibitem{Verresen2021}
Ruben {Verresen}, Nathanan {Tantivasadakarn}, and Ashvin {Vishwanath}.
\newblock ``{Efficiently preparing Schr{\"o}dinger's cat, fractons and non-Abelian topological order in quantum devices}''~(2021).
\newblock  \href{http://arxiv.org/abs/2112.03061}{arXiv:2112.03061}.

\bibitem{Bravyi2022}
Sergey {Bravyi}, Isaac {Kim}, Alexander {Kliesch}, and Robert {Koenig}.
\newblock ``{Adaptive constant-depth circuits for manipulating non-abelian anyons}''~(2022).
\newblock  \href{http://arxiv.org/abs/2205.01933}{arXiv:2205.01933}.

\bibitem{Tantivasadakarn2023nonabelian}
Nathanan Tantivasadakarn, Ruben Verresen, and Ashvin Vishwanath.
\newblock ``Shortest route to non-abelian topological order on a quantum processor''.
\newblock \href{https://dx.doi.org/10.1103/PhysRevLett.131.060405}{Phys. Rev. Lett. {\bf 131}, 060405}~(2023).

\bibitem{Li2023}
Yabo Li, Hiroki Sukeno, Aswin~Parayil Mana, Hendrik~Poulsen Nautrup, and Tzu-Chieh Wei.
\newblock ``Symmetry-enriched topological order from partially gauging symmetry-protected topologically ordered states assisted by measurements''.
\newblock \href{https://dx.doi.org/10.1103/PhysRevB.108.115144}{Phys. Rev. B {\bf 108}, 115144}~(2023).

\bibitem{Williamson2024}
Dominic~J. {Williamson} and Theodore~J. {Yoder}.
\newblock ``{Low-overhead fault-tolerant quantum computation by gauging logical operators}''~(2024).
\newblock  \href{http://arxiv.org/abs/2410.02213}{arXiv:2410.02213}.

\bibitem{Mochon2004}
Carlos Mochon.
\newblock ``Anyon computers with smaller groups''.
\newblock \href{https://dx.doi.org/10.1103/PhysRevA.69.032306}{Phys. Rev. A {\bf 69}, 032306}~(2004).

\bibitem{Cui2015}
Shawn~X. {Cui}, Seung-Moon {Hong}, and Zhenghan {Wang}.
\newblock ``{Universal quantum computation with weakly integral anyons}''.
\newblock \href{https://dx.doi.org/10.1007/s11128-015-1016-y}{Quantum Information Processing {\bf 14}, 2687--2727}~(2015).
\newblock  \href{http://arxiv.org/abs/1401.7096}{arXiv:1401.7096}.

\bibitem{Laubscher2019}
Katharina Laubscher, Daniel Loss, and James~R. Wootton.
\newblock ``Universal quantum computation in the surface code using non-abelian islands''.
\newblock \href{https://dx.doi.org/10.1103/PhysRevA.100.012338}{Phys. Rev. A {\bf 100}, 012338}~(2019).

\bibitem{Ren2023}
Yuanjie {Ren} and Peter {Shor}.
\newblock ``{Topological quantum computation assisted by phase transitions}''~(2023).
\newblock  \href{http://arxiv.org/abs/2311.00103}{arXiv:2311.00103}.

\bibitem{Farinholt2014}
J~M Farinholt.
\newblock ``An ideal characterization of the clifford operators''.
\newblock \href{https://dx.doi.org/10.1088/1751-8113/47/30/305303}{Journal of Physics A: Mathematical and Theoretical {\bf 47}, 305303}~(2014).

\bibitem{Cowtan2022}
Alexander {Cowtan}.
\newblock ``{Qudit lattice surgery}''~(2022).
\newblock  \href{http://arxiv.org/abs/2204.13228}{arXiv:2204.13228}.

\bibitem{Gidney2024Ybasis}
Craig {Gidney}.
\newblock ``{Inplace Access to the Surface Code Y Basis}''.
\newblock \href{https://dx.doi.org/10.22331/q-2024-04-08-1310}{Quantum {\bf 8}, 1310}~(2024).
\newblock  \href{http://arxiv.org/abs/2302.07395}{arXiv:2302.07395}.

\bibitem{Bombin2008}
H.~Bombin and M.~A. Martin-Delgado.
\newblock ``Family of non-abelian kitaev models on a lattice: Topological condensation and confinement''.
\newblock \href{https://dx.doi.org/10.1103/PhysRevB.78.115421}{Phys. Rev. B {\bf 78}, 115421}~(2008).

\bibitem{Bhardwaj2023Club}
Lakshya {Bhardwaj}, Lea~E. {Bottini}, Daniel {Pajer}, and Sakura {Schafer-Nameki}.
\newblock ``{The Club Sandwich: Gapless Phases and Phase Transitions with Non-Invertible Symmetries}''~(2023).
\newblock  \href{http://arxiv.org/abs/2312.17322}{arXiv:2312.17322}.

\bibitem{Bhardwaj2024Hasse}
Lakshya {Bhardwaj}, Daniel {Pajer}, Sakura {Schafer-Nameki}, and Alison {Warman}.
\newblock ``{Hasse Diagrams for Gapless SPT and SSB Phases with Non-Invertible Symmetries}''~(2024).
\newblock  \href{http://arxiv.org/abs/2403.00905}{arXiv:2403.00905}.

\bibitem{Ren2024}
Yuanjie {Ren}, Nathanan {Tantivasadakarn}, and Dominic~J. {Williamson}.
\newblock ``{Efficient Preparation of Solvable Anyons with Adaptive Quantum Circuits}''~(2024).
\newblock  \href{http://arxiv.org/abs/2411.04985}{arXiv:2411.04985}.

\bibitem{Ellison2022}
Tyler~D. Ellison, Yu-An Chen, Arpit Dua, Wilbur Shirley, Nathanan Tantivasadakarn, and Dominic~J. Williamson.
\newblock ``Pauli stabilizer models of twisted quantum doubles''.
\newblock \href{https://dx.doi.org/10.1103/PRXQuantum.3.010353}{PRX Quantum {\bf 3}, 010353}~(2022).

\bibitem{Iqbal2024D4}
Mohsin Iqbal, Nathanan Tantivasadakarn, Ruben Verresen, Sara~L. Campbell, Joan~M. Dreiling, Caroline Figgatt, John~P. Gaebler, Jacob Johansen, Michael Mills, Steven~A. Moses, Juan~M. Pino, Anthony Ransford, Mary Rowe, Peter Siegfried, Russell~P. Stutz, Michael Foss-Feig, Ashvin Vishwanath, and Henrik Dreyer.
\newblock ``Non-abelian topological order and anyons on a trapped-ion processor''.
\newblock \href{https://dx.doi.org/10.1038/s41586-023-06934-4}{Nature {\bf 626}, 505--511}~(2024).

\bibitem{Wootton2014}
James~R. Wootton, Jan Burri, Sofyan Iblisdir, and Daniel Loss.
\newblock ``Error correction for non-abelian topological quantum computation''.
\newblock \href{https://dx.doi.org/10.1103/PhysRevX.4.011051}{Phys. Rev. X {\bf 4}, 011051}~(2014).

\bibitem{Cui2020kitaevsquantum}
Shawn~X. Cui, Dawei Ding, Xizhi Han, Geoffrey Penington, Daniel Ranard, Brandon~C. Rayhaun, and Zhou Shangnan.
\newblock ``Kitaev's quantum double model as an error correcting code''.
\newblock \href{https://dx.doi.org/10.22331/q-2020-09-24-331}{{Quantum} {\bf 4}, 331}~(2020).

\bibitem{Margarita2025}
Margarita {Davydova}, Andreas {Bauer}, Julio~C. {Magdalena de la Fuente}, Mark {Webster}, Dominic~J. {Williamson}, and Benjamin~J. {Brown}.
\newblock ``{Universal fault tolerant quantum computation in 2D without getting tied in knots}''~(2025).
\newblock  \href{http://arxiv.org/abs/2503.15751}{arXiv:2503.15751}.

\bibitem{Frohlich2006}
J{\"u}rg Fr{\"o}hlich, J{\"u}rgen Fuchs, Ingo Runkel, and Christoph Schweigert.
\newblock ``Correspondences of ribbon categories''.
\newblock \href{https://dx.doi.org/https://doi.org/10.1016/j.aim.2005.04.007}{Advances in Mathematics {\bf 199}, 192--329}~(2006).

\bibitem{Chatterjee2023Holo}
Arkya Chatterjee and Xiao-Gang Wen.
\newblock ``Holographic theory for continuous phase transitions: Emergence and symmetry protection of gaplessness''.
\newblock \href{https://dx.doi.org/10.1103/PhysRevB.108.075105}{Phys. Rev. B {\bf 108}, 075105}~(2023).

\end{thebibliography}

\end{document}